\documentclass[10pt,journal,twocolumn]{IEEEtran}
\usepackage{ifpdf}
 \ifpdf
 \else
 \fi
\ifCLASSINFOpdf
   \usepackage[pdftex]{graphicx}
\else
   \usepackage[dvips]{graphicx}
\fi
\usepackage[cmex10]{amsmath}

\usepackage[tight,footnotesize]{subfigure}
\usepackage{define-notations,amssymb}
\usepackage{algorithmic,algorithm,cite,color}
\newtheorem{lemma}{Lemma}
\newtheorem{theorem}{Theorem}

\begin{document}
\title{On the Optimum Energy Efficiency for Flat-fading Channels
with Rate-dependent Circuit Power}

\author{
Tao Wang,~\IEEEmembership{Senior Member, IEEE},
Luc Vandendorpe,~\IEEEmembership{Fellow, IEEE}
\thanks{
T. Wang is with Key Laboratory of Specialty Fiber Optics
and Optical Access Networks,
School of Communication \& Information Engineering,
Shanghai University, 200072 Shanghai, P. R. China (email: t.wang@ieee.org
or twang@shu.edu.cn).
He was with ICTEAM Institute, Universit\'e Catholique de Louvain,
1348 Louvain-la-Neuve, Belgium.

L. Vandendorpe is with ICTEAM Institute,
Universit\'e Catholique de Louvain, 1348 Louvain-la-Neuve, Belgium
(email: luc.vandendorpe@uclouvain.be).

This research is supported by The Program for Professor of
Special Appointment (Eastern Scholar) at Shanghai Institutions of Higher Learning,
Innovation Program of Shanghai Municipal Education Commission (14ZZ096),
and Innovation Fund of Shanghai University (K.10-0107-13-001).
It is also supported by the IAP project BESTCOM,
the European Commission in the framework of the FP7 Network of Excellence
in Wireless COMmunications NEWCOM\# (Grant agreement no. 318306).
This paper was presented in part
at the International Conference on Wireless Communications and Signal Processing, Hangzhou, China, 2013.}}

\markboth{To appear in IEEE Transactions on Communications}
{On the Optimum Energy Efficiency for Flat-fading Channels
with Rate-dependent Circuit Power}

\maketitle

\begin{abstract}
This paper investigates the optimum energy efficiency (EE)
and the corresponding spectral efficiency (SE) for a communication link
operating over a flat-fading channel.
The EE is evaluated by the total energy consumption
for transmitting per message bit.
Three channel cases are considered, namely static channel with channel state
information available at transmitter (CSIT),
fast-varying (FV) channel with channel distribution information available at transmitter (CDIT), and FV channel with CSIT.
The link's circuit power is modeled as
$\Pcst + \CoeffPrate\Prate(R)$ Watt,
where $\Pcst>0$ and $\CoeffPrate\geq0$ are two constants
and $\Prate(R)$ is a general increasing and convex function
of the transmission rate $R\geq0$.
For all the three channel cases, the tradeoff between the EE and SE is
studied.
It is shown that the EE improves strictly
as the SE increases from $0$ to the optimum SE,
and then strictly degrades as the SE increases beyond the optimum SE.
The impact of $\CoeffPrate$, $\Pcst$
and other system parameters on the optimum EE and corresponding SE
is investigated to obtain insight.
Some of the important and interesting results
for all the channel cases include:
(1) when $\CoeffPrate$ increases the SE corresponding to the optimum EE
should keep unchanged if $\Prate(R)=R$,
but reduced if $\Prate(R)$ is strictly convex of $R$;
(2) when the rate-independent circuit power $\Pcst$ increases,
the SE corresponding to the optimum EE has to be increased.
A polynomial-complexity algorithm is developed with the bisection method
to find the optimum SE.
The insight is corroborated and the optimum EE for
the three cases are compared by simulation results.
\end{abstract}


\section{Introduction}

Energy-efficient communication and signal processing techniques
play important roles in applications where devices are powered
by batteries \cite{Li11,Chen11,Wang09,Wang10,Wang12,Wang12CRB,Wang13WCSP}.
For a communication system, its energy efficiency (EE)
can be evaluated by either the total energy consumption
for transmitting per message bit (TEPB),
or the number of message bits transmitted with
per-Joule total energy consumption (NBPE).
A higher EE is represented by either a smaller TEPB or a greater NBPE.
{Note that due to the scarcity of spectral resource,
there already existed traditional and intensive research on
increasing spectral efficiency (SE) as an important goal
in the field of wireless communications.
Therefore, it becomes a very important research topic to study
the relationship between the optimum EE and the corresponding SE,
as well as the impact of system parameters on them
for wireless communication systems.
}

Early works studying the EE of communication systems
only considered transmission energy but ignored circuit energy consumption.
For instance, approximate expressions of per message bit transmission energy
were derived in \cite{Verdu02} as a function of the spectral efficiency
for flat-fading channels in wideband regime,
and some strategies to reduce the TEPB were discussed
in \cite{Gamal02,Schurgers02}.
{In these works, only the transmission energy consumed for radiating
radio-frequency signals was taken in account while
the circuit energy consumption was neglected,
which makes senses for long-distance communication related application scenarios.
The major finding is that the SE has to be reduced 
to improve the EE when only the transmission energy is considered,
i.e., the SE and EE are contradictory performance metrics
since improving one leads to degradation of another one.
}

For the high-EE design of short-distance communication systems,
which have many promising applications and
thus attracted much research interest,
the circuit energy consumption however cannot be ignored \cite{Shih2001}.
For instance, data transmission within a wireless body area network
is mainly over short distance, which leads to small transmission energy
consumption comparable to the circuit energy consumption \cite{WBAN}.
In such a case, the circuit energy must be taken into account.
In view of the above fact, the circuit energy was taken into account
to optimize the EE of communication systems in recent works.
For instance, modulation schemes were optimized in \cite{Cui05}
for communication links operating over flat-fading channels,
and link adaptation algorithms were developed
in \cite{Miao10,Ish10,Xiong11,Miao12,Xiong12,Wang13ICCC} for multi-carrier systems transmitting over frequency-selective channels.
General frameworks for energy efficiency optimization
were proposed in \cite{Ish12,Chong12}.
In \cite{Cui05,Miao10,Miao12,Ish12,Chong12,Wang13ICCC}, the circuit power
was assumed to remain fixed independently of the bit transmission rate.
In \cite{Ish10,Xiong11,Xiong12}, the circuit power was assumed to be
linear with the transmission rate.
In general, the circuit power is an increasing function of
the transmission rate, since a greater bit rate indicates
that a bigger codebook is used which usually incurs
higher power for encoding and decoding on baseband circuit boards
\cite{ranpara1999low,zhang2010efficient}.
{Note that static channels with channel state information
at the transmitter (CSIT) were studied in \cite{Cui05,Miao10,Ish10,Xiong11,Miao12,Wang13ICCC},
while both static and fast-varying channels with
CSIT were studied in \cite{Ish12,Chong12}.}
{The major finding in these works is that
when taking into account the circuit energy consumption,
the relationship between the SE and EE is
{\it fundamentally different} from that when the circuit energy is ignored.
In particular, the EE usually first improves then degrades
as the SE increases from zero \cite{Wang13ICCC}.
}

In this paper, we study the optimum EE and the corresponding SE for
flat fading channels with rate-dependent circuit power
in a more general form than those studied previously.
Even though the flat-fading channel model seems simple,
it deserves research effort due to the following reasons.
First, it has been widely used in practice especially for low-power
applications, e.g., in wireless sensor networks
where highly energy-efficient transmission is needed.
Second, there exist different cases when using the flat-fading channel,
which depend on the condition of channel variation
and availability of channel knowledge.
For these cases, the optimum EE and the corresponding SE performance
deserve much attention and need to be thoroughly investigated.
Motivated by the above fact, we consider three different
cases for using the flat-fading channel, namely
\begin{enumerate}
\item
Case 1: static channel with CSIT;
\item
Case 2: fast-varying channel with channel distribution information at transmitter (CDIT);
\item
Case 3: fast-varying channel with CSIT.
\end{enumerate}

For each case listed above,
we model the link's total power consumption as the sum of
the power amplifier's power consumption and circuit power.
The circuit power is modeled as the sum of a constant
power and a rate-dependent part which is a general increasing
and convex function of the transmission rate.
This circuit-power model
is more general than those studied in the literature
since either fixed circuit power or the circuit power as
a linear function of transmission rate was studied previously.
We formulate the EE-SE function and make EE-SE tradeoff study.
The impact of system parameters on the optimum EE and SE
is then studied.
In particular, insight is obtained from the theoretical
analysis, which may help practical system design to improve its EE.
A polynomial-complexity algorithm is developed with the
bisection method to find the optimum EE and SE.
Finally, we show simulation results to corroborate
the insight obtained from the theoretical analysis
and compare the optimum EEs for the different channel cases.

The rest of this paper is organized as follows.
The system models are described in the next Section.
After that, the EE-SE function is formulated and
the EE-SE tradeoff analysis is made in Section III.
The impact of system parameters on
the optimum EE and SE is investigated in Section IV.
The algorithm is developed in Section V, and
simulation results are shown in Section VI to illustrate
the obtained insight. Some conclusions are made in Section VII.

Notations: $\Avg{{x}}{f(x)}$ represents the ensemble average of
the function $f(x)$ over the probability density function of
the random variable $x$.
$y'(x)$ and $y''(x)$ indicate the first-order and second-order
derivatives of $y(x)$ with respect to $x$, respectively.

\section{System models}

Consider a point-to-point communication link
transmitting over a flat-fading channel using a bandwidth $\BW$ Hz.
The baseband channel model is formulated as
\begin{align}
y = h x + n
\end{align}
where $x$ is the complex symbol emitted by the transmitter,
and $y$ represents the corresponding symbol received at the receiver's
baseband processor. $h$ is the channel coefficient.
$n$ is the sum of additive white Gaussian noise
and the cochannel interference.
We assume $n$ is a random variable with circularly symmetric
complex Gaussian distribution with zero mean and variance $\Nvar$,
which keeps invariant during data transmission.
$h$ and $\Nvar$ are assumed to be known by the receiver.

The link's total power consumption is modeled as the sum of two parts:
the power consumed by the transmitter's power amplifier
for emitting coded symbols and circuit power.
Specifically, the circuit power is modeled as
$\Pcst + \CoeffPrate\Prate(R)$ Watt,
where $\Pcst>0$ and $\CoeffPrate\geq0$ are two constants
and $R$ represents the transmission rate in the unit of bits/second.
$\Pcst$ represents the rate-independent circuit power
which models the sum power of filters, low-noise-amplifiers,
mixers, synthesizers, etc.
$\CoeffPrate\Prate(R)$ models the rate-dependent circuit power,
e.g. that consumed by channel encoder and decoder.
We assume $\Prate(R)$ satisfies that
\begin{enumerate}
\item
$\Prate(0)=0$, i.e., the rate-dependent circuit power is zero
when $R = 0$;
\item
$\Prate(R)$ is differentiable, strictly increasing
and (not necessarily strictly) convex of $R\geq0$.
\end{enumerate}

Note that the rate-dependent circuit power models
studied in the literature, e.g., \cite{Cui05,Miao10,Ish10,Xiong11}
are special cases of the model assumed above.
Specifically,
when the rate-dependent circuit power is negligible
as in \cite{Cui05,Miao10}, we can simply set $\CoeffPrate=0$.
When the rate-dependent circuit power increases linearly with respect
to the rate as studied in \cite{Ish10,Xiong11},
we can set $\Prate(R)=R$ and $\CoeffPrate$
as the increasing rate.
Moreover, the model is also applicable for the links
where the rate-dependent circuit power is strictly convex of the rate
as will be studied later in this paper.

{
Define $G = |h|^2$ as the instantaneous channel power gain.
Three different scenarios for using the communication link
will be considered as follows:
\begin{itemize}
\item
Case 1 (static channel with CSIT):
the channel keeps invariant with CSI available at the transmitter,
and $G$ is known by the transmitter at the beginning of the transmission.

\item
Case 2 (FV channel with CDIT):
the channel varies during the data transmission and
the probability density function (pdf) of $G$ is known a priori by the transmitter.

\item
Case 3 (FV channel with CSIT):
the channel varies during the data transmission and
$G$ is known during the transmission.
\end{itemize}
}

For Case 1 and Case 2, suppose the average power of
transmitted symbols (referred to as transmission power hereafter)
is $p$ Watt, i.e., $\Avg{x}{|x|^2} = P$.
Assume the optimum codebook is used, the maximum SE can be evaluated as:
\begin{align}
\SE(P) = \left\{\begin{array}{ll}
                    \log_2\big(1 + G\frac{P}{\Nvar}\big)           &{\rm\hspace{0.1cm} for\;Case\;1}; \\
                    \Avg{G}{\log_2\big(1 + G\frac{P}{\Nvar}\big)}  &{\rm\hspace{0.1cm} for\;Case\;2}.
                \end{array}
         \right.
\end{align}
in the unit of bits/second/Hz.
For both cases, the average sum power is
\begin{align}
\CoeffPrate\Prate(\BW\SE(P)) + \frac{P}{\PAEff} + \Pcst {\hspace{0.3cm}\rm (Watt)}
\end{align}
where $\PAEff$ represents the efficiency of the power amplifier.

For Case 3, the transmission power can be adapted according to CSI.
Suppose the transmitter uses $P(G)$ as the transmission power
when the channel power gain is $G$.
Note that any nonnegative function of $G$ can be assigned to $P(G)$
as a feasible power-allocation strategy,
denoted by $\Pstrat = \{P(G)|\forall\;G\geq 0\}$ hereafter.
Obviously, the set of all feasible strategies is simply
the set of all nonnegative functions, denoted by $\Pspace$ hereafter.
Assume the optimum codebook is used,
the SE corresponding to using $\Pstrat$ is equal to
\begin{align}
\SE(\Pstrat) = \Avg{G}{\log_2\big(1 + G\frac{P(G)}{\Nvar}\big)}
\label{eq:maxSE-CSIT-FV}
\end{align}
in the unit of bits/second/Hz. The average sum power is
\begin{align}
\CoeffPrate\Prate(\BW\SE(\Pstrat)) + \frac{\Avg{G}{P(G)}}{\xi} + \Pcst
\hspace{0.3cm}{\rm (Watt)}.
\end{align}

\section{EE-SE tradeoff formulation and analysis}

{
In this paper, the EE is evaluated as the TEPB as in \cite{Cui05}.
Obviously, a smaller TEPB indicates a better EE.
For the single-user case as studied in this paper,
it is also equivalent to express it as the NBPE.
For multi-user scenarios as studied in \cite{Miao12},
it might be more convenient to use the NBPE to evaluate the EE.
}

To facilitate the EE-SE tradeoff analysis,
we first formulate in Section \ref{sec:formulation} for each case
the EE-SE function $\EE(\SE)$, defined as the EE corresponding to a given SE $\SE$.
Denote the optimum EE as $\optEE$, i.e.,
\begin{align}
\optEE = \min_{\SE\geq0}\EE(\SE),
\end{align}
and the corresponding optimum SE as $\optSE$, i.e,
\begin{align}
\optSE = \arg\min_{\SE\geq0}\EE(\SE).
\end{align}

Note that for each case under consideration,
$\optSE>0$ must hold because when $\SE=0$, the $\EE$ is $+\infty$
due to the existence of $\Pcst>0$.
In Section \ref{sec:analysis} we will make theoretical analysis
and show geometric interpretation to study properties of
$\EE(\SE)$, $\optEE$ and $\optSE$, which unveils the EE-SE tradeoff.

\subsection{Formulation of the EE-SE function}\label{sec:formulation}

To formulate $\EE(\SE)$, we first derive $\PowToNvar(\SE)$,
defined as the ratio of the minimum transmission power
required to achieve $\SE$ to $\Nvar$.
It can readily be shown that
\begin{itemize}
\item
For Case 1:
\begin{align}
\PowToNvar(\SE) = \min_{p\geq0: \log_2(1 + Gp) \geq \SE}  \hspace{0.2cm}  p
= \frac{1}{G}(2^\SE - 1)                        \label{eq:PowToNvar-SE-case1}
\end{align}

\item
For Case 2:
\begin{align}
\PowToNvar(\SE) = \min_{p\geq0}  &\hspace{0.2cm}  p
\label{eq:PowToNvar-SE-case2}\\
{\rm s.t.}  &\hspace{0.2cm} \Avg{G}{\log_2(1 + Gp)} \geq \SE.  \nonumber
\end{align}

\item
For Case 3:
\begin{align}
\PowToNvar(\SE) = \min_{\{p(G)|\forall\;G\geq0\}\in\Pspace} &\hspace{0.2cm}  \Avg{G}{p(G)}   \label{eq:PowToNvar-SE-case3} \\
{\rm s.t.}  &\hspace{0.2cm} \Avg{G}{\log_2(1 + Gp(G))} \geq \SE.  \nonumber
\end{align}
\end{itemize}

Note that $\PowToNvar(\SE)$ for Case 1 depends only on $G$
while $\PowToNvar(\SE)$ for the other two cases relies only on the pdf of $G$.
According to the above formulas,
the minimum total power required to achieve $\SE$ for each case
can be evaluated with a unified expression as:
\begin{align}
P(\SE) = \CoeffPrate\Prate(\BW\SE)
         + \frac{\Nvar}{\PAEff}\PowToNvar(\SE)
         + \Pcst   {\rm\hspace{0.2cm}(Watt)},
\end{align}
and the EE given $\SE$ is evaluated by
\begin{align}
\EE(\SE) = \frac{P(\SE)}{\BW\SE}
= \frac{\CoeffPrate\Prate(\BW\SE)
         + \frac{\Nvar}{\PAEff}\PowToNvar(\SE)
         + \Pcst}{\BW\SE}   {\rm\hspace{0.2cm}(Joule/bit)}.
\end{align}

\subsection{EE-SE tradeoff analysis}\label{sec:analysis}

To facilitate the tradeoff analysis,
we first show the following property of $\PowToNvar(\SE)$:
\begin{lemma}\label{lemma:PowToNvar-SE-convex}
For each channel case under consideration,
$\PowToNvar(\SE)$ is strictly increasing and strictly convex of $\SE\geq 0$.
\end{lemma}
\begin{IEEEproof}
See the Appendix.
\end{IEEEproof}

We then show an important property of $\EE(\SE)$ as follows:

%=================== quasiconvexity: most important ===================
\begin{lemma}\label{lemma:EE-SE-quasiconvex}
For each channel case and any $\Prate(R)$ satisfying the assumptions made in Section II,
$\EE(\SE)$ is a strictly quasiconvex function of $\SE$.
\end{lemma}
\begin{IEEEproof}
See the Appendix.
\end{IEEEproof}
%==================================================

%====================== Properties ===================

According to the strict quasiconvexity of $\EE(\SE)$,
\begin{align}
\max\{\EE(\SE_1), \EE(\SE_2)\} > \EE(\SE)   \label{eq:quasiconvex-ineq}
\end{align}
holds $\forall\;\SE\in(\SE_1,\SE_2)$ \cite{Quasi-convex,Convex-opt}.
The strict quasiconvexity is a key feature for $\EE(\SE)$,
based on which we will prove properties of $\optSE$ and $\EE(\SE)$
in the following.
To facilitate description, we first derive the derivative
of $\EE(\SE)$ with respect to $\SE$ as follows:
\begin{align}
\EE'(\SE) = \frac{1}{\BW\SE}\left[P'(\SE) - \frac{P(\SE)}{\SE}\right].
%\Der{\EE(\SE)}{\SE}
%= \frac{1}{\BW\SE}\left[\Der{P(\SE)}{\SE} - \frac{P(\SE)}{\SE}\right].
\label{eq:Der-EE}  \end{align}

\begin{theorem}\label{theorem:EE-SE-tradeoff}
For each channel case, the following properties are satisfied:
\begin{enumerate}
\item
There exists a unique $\SE^\star$ and it satisfies
\begin{align}
P'(\optSE) = \frac{P(\optSE)}{\SE} \label{eq:optSE-cond}
\end{align}
\item
$\EE(\SE)$ is strictly decreasing with $\SE\in(0,\SE^\star]$ and
\begin{align}
\forall\;\SE\in(0,\SE^\star), P'(\SE) < \frac{P(\SE)}{\SE}.   \label{eq:leftSE-cond}
\end{align}
\item
$\EE(\SE)$ is strictly increasing with $\SE\in[\SE^\star,+\infty)$ and
\begin{align}
\SE\in(\SE^\star,+\infty), P'(\SE) > \frac{P(\SE)}{\SE}.    \label{eq:rightSE-cond}
\end{align}
\end{enumerate}
\end{theorem}
\begin{IEEEproof}
See the Appendix.
It is interesting to note that these properties
are derived by solely using the strict quasiconvexity of $\EE(\SE)$,
without resorting to the derivative of $\EE(\SE)$ with respect to $\SE$
as in \cite{Miao10,Xiong11,Xiong12}
\footnote{Note that this theorem as well as Algorithm 1 given later
are applicable for three cases of point-to-point flat-fading channel models
which can be either invariant or time-varying.
Theorem 1 and Algorithm 1 in \cite{Xiong12} which are similar to 
them are applicable for time-invariant multi-user frequency-selective channels.  
}.

\end{IEEEproof}

%====================== Geometric interpretations ===================
\begin{figure}
  \centering
  \subfigure[Illustration of $\EE(\SE_l)$ and $\EE(\SE_l)$ for $\SE_l\leq \optSE$]{
     \includegraphics[width=3.5in]{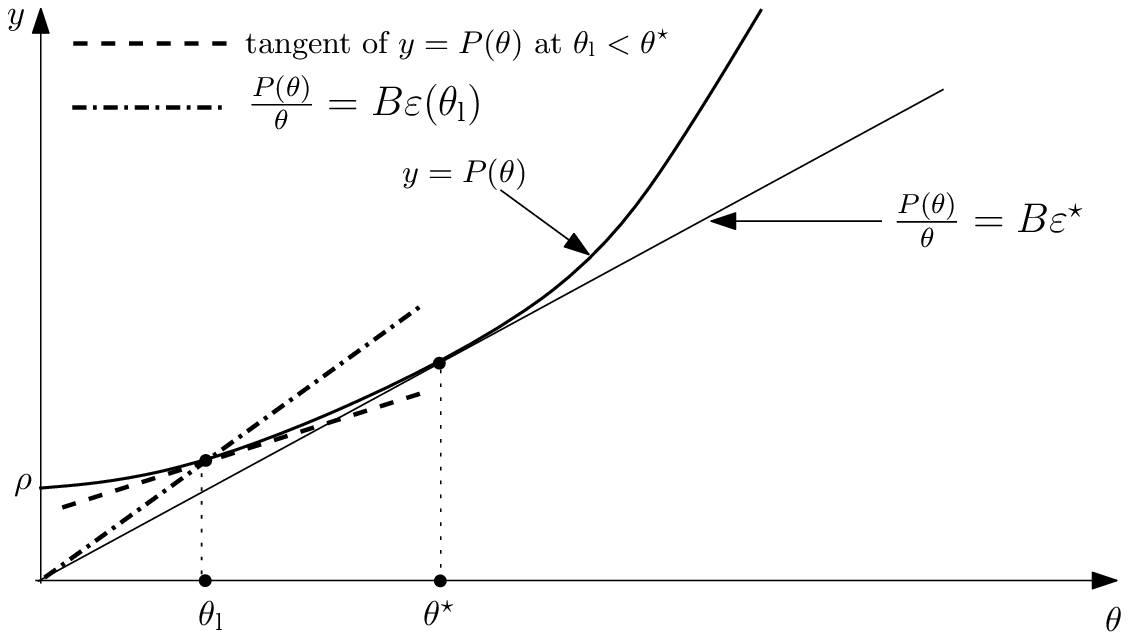}}
  \subfigure[Illustration of $\EE(\SE_u)$ and $\EE(\SE_u)$ for $\SE_u\geq \optSE$]{
     \includegraphics[width=3.5in]{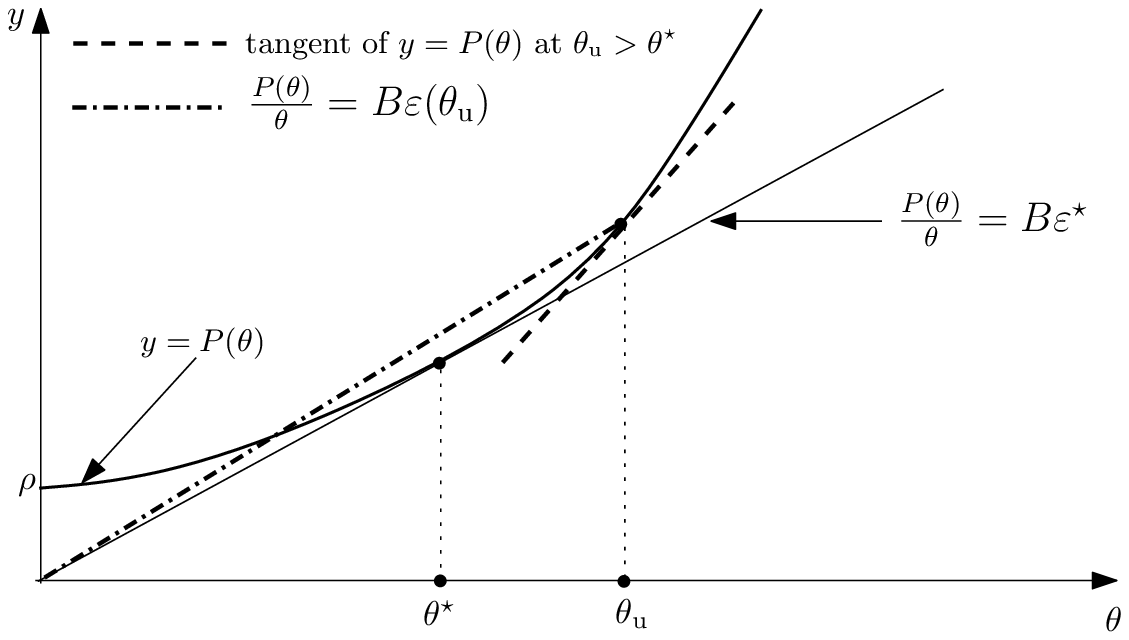}}
  \caption{Geometric interpretation for the EE-SE tradeoff analysis.}  \label{fig:geometric-interpret}
\end{figure}

Straightforward geometric interpretations can be presented
to intuitively explain the monotonic properties of $\EE(\SE)$
specified in Theorem \ref{theorem:EE-SE-tradeoff}.
Specifically, we can plot the line
$\mathcal{L}_1 = \{\Coord(\SE) = (\SE,P(\SE))|\SE\geq0\}$
over the two-dimensional plane of coordinates $(\SE,y)$
as shown in Fig. \ref{fig:geometric-interpret}.
Note that $\BW\EE(\SE)=\frac{P(\SE)}{\SE}$ is equal to
the slope of the origin-to-$\Coord(\SE)$ line,
and $P'(\SE)$ is equal to the slope of
the tangent line of $y=P(\SE)$ at $\SE$.

It can be seen from Fig. \ref{fig:geometric-interpret}
that as $\Coord(\SE)$ moves away from the origin along the line $\mathcal{L}_1$,
the slope of the origin-to-$\Coord(\SE)$ line
first strictly reduces and then strictly increases,
meaning that $\EE(\SE)$ first strictly reduces and then strictly increases.
The reason behind this observation is the strict convexity of $P(\SE)$.
At $\SE=\optSE$ where $\EE(\SE)$ is minimized
(i.e., the slope of the origin-to-$\Coord(\SE)$ line is the smallest),
the origin-to-$\Coord(\SE)$ line coincides with the tangent line of $y=P(\SE)$,
indicating that \eqref{eq:optSE-cond} indeed holds.
It can be seen that $\optSE$ is unique due to the strict convexity of $P(\SE)$.
For any $\SE\in(0,\optSE)$, it can be seen from Fig. \ref{fig:geometric-interpret}.a
that the origin-to-$\Coord(\SE)$ line is steeper than the tangent line
of $y=P(\SE)$, indicating that \eqref{eq:leftSE-cond} indeed holds.
For any $\SE\in(\optSE,+\infty)$, it can be seen from Fig. \ref{fig:geometric-interpret}.b
that the tangent line of $y=P(\SE)$ is steeper than
the origin-to-$\Coord(\SE)$ line, indicating that
\eqref{eq:rightSE-cond} indeed holds.
{Note that a similar geometric interpretation
was exhibited in \cite{Ish12} to derive optimality conditions
for solving the problem of maximizing $\frac{y_1(x)}{y_2(x)}$
where $y_1(x)$ is concave and $y_2(x)$ is a nonnegative
convex function. }

\begin{figure}
  \centering
     \includegraphics[width=3.5in]{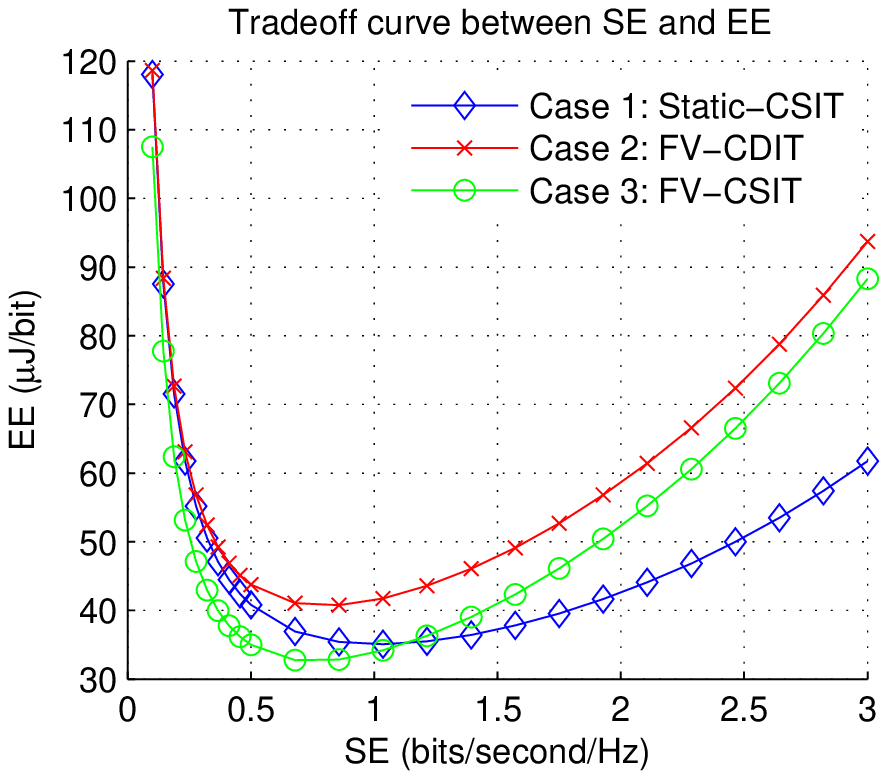}
     \caption{Illustration of $\EE(\SE)$ when $\Pcst=100$ mW, $\Nvar=-80$ dBW,
              $\BW=10$ KHz, $\PAEff=0.4$, $\CoeffPrate=8\times10^{-8}$, and $\Prate(R)=R$.
              For Case 1, $G=-70$ dB. For Case 2 and Case 3, $h$ follows
              rayleigh distribution with average channel power gain
              equal to $-70$ dB. }
  \label{fig:illustrate-EE}
\end{figure}

The fundamental insight behind Theorem \ref{theorem:EE-SE-tradeoff}
is that increasing $\SE$ is favorable for improving $\EE$
when $\SE\leq \optSE$, while $\SE$ has to be sacrificed
for better $\EE$ when $\SE\geq \optSE$.
To illustrate the shape of $\EE(\SE)$,
the $\EE(\SE)$ is plotted in Fig. \ref{fig:illustrate-EE}
for the three channel cases over a typical flat-fading channel.
The monotonic properties of $\EE(\SE)$ claimed
in Theorem \ref{theorem:EE-SE-tradeoff} can be seen clearly.

\section{Impact of system parameters on $\optEE$ and $\optSE$}\label{sec:impact}

Collect system parameters for Case 1 into the set
$\chi=\{G,\Nvar,\CoeffPrate,\Pcst\}$,
and those for Cases 2 and 3 into the set
$\chi=\{\Nvar,\CoeffPrate,\Pcst\}$.
Denote the $\optEE$ and $\optSE$ corresponding to a given $\chi$
as $\optEE(\chi)$ and $\optSE(\chi)$, respectively.
In the following, we will study the impact of system
parameters on $\optEE(\chi)$ and $\optSE(\chi)$.

%\subsection{General results about the system parameter's impact}

We first present some preliminary rules
that will play important roles later:

\begin{lemma}\label{lemma:judge-SE}
For each case under consideration, given any $\SE\geq0$,
it must satisfy
\begin{align}
\SE \hspace{0.1cm}\left\{\begin{array}{ll}
                              <\optSE(\chi)
                              &{\rm iff}\hspace{0.2cm} \Gamma_\chi(\SE) < 0\\
                              = \optSE(\chi)
                              &{\rm iff}\hspace{0.2cm} \Gamma_\chi(\SE) = 0\\
                              > \optSE(\chi)
                              &{\rm iff}\hspace{0.2cm} \Gamma_\chi(\SE) > 0
                          \end{array} \right.        \label{eq:judge-SE}
\end{align}
where
\begin{align}
\Gamma_\chi(\SE) = \CoeffPrate g(\BW\SE) + \frac{\Nvar}{\PAEff}f(\SE) - \Pcst,
\end{align}
and
\begin{align}
g(R)   = R\Prate'(R) - \Prate(R),\;\;
f(\SE) = \SE\PowToNvar'(\SE) - \PowToNvar(\SE). \label{eq:expression-fSE-gSE}
\end{align}

Moreover, the following claims are true:
\begin{enumerate}
\item
$g(0)=0$. If $\Prate(R)$ is strictly convex of $R\geq0$,
$g(R)$ is strictly increasing of $R\geq0$,
while $g(R)=0$, $\forall\;R\geq0$ if $\Prate(R) = R$.
\item
$f(0)=0$ and $f(\SE)$ is strictly increasing of $\SE\geq0$.
\end{enumerate}
\end{lemma}
\begin{IEEEproof}
See the Appendix.
\end{IEEEproof}

\begin{figure}
  \centering
     \includegraphics[width=3.5in]{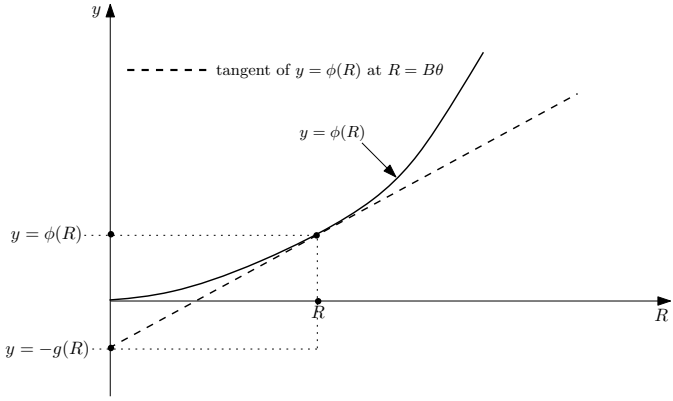}
  \caption{Geometric interpretation for the properties of $g(R)$
  when $\Prate(R)$ is strictly convex of $R\geq0$.}  \label{fig:g-convex-R}
\end{figure}

Straightforward geometric interpretation can be given
to corroborate the above properties of $g(R)$ intuitively
(those for $f(\SE)$ can be interpreted in a similar way
and are thus omitted here).
Specifically, we can plot the line
$\mathcal{L}_2 = \{\Coord(R) = (R,\Prate(R))|\forall\;R\geq0\}$
over the two-dimensional plane of coordinates $(R,y)$
when $\Prate(R)$ is strictly convex of $R\geq0$
as shown in Fig. \ref{fig:g-convex-R}.
Most interestingly, $-g(R) = \Prate(R) - R\Prate'(R)$
is equal to the y-coordinate of the intersection
between the line $R=0$ and the tangent line of $y=\Prate(R)$
drawn at the coordinate $\Coord(R)$.
When $\Coord(R)$ moves away from the origin
along the line $\mathcal{L}_2$,
$-g(R)$ strictly decreases, which means that
$g(R)$ indeed strictly increases.
When $\Prate(R)=R$,
it can be seen that $g(R)$ is fixed as $0$ for any $R\geq0$.

%=================== impact-kappa =====================
Based on the above lemma, we first show the impact of $\CoeffPrate$ on
$\optSE(\chi)$ and $\optEE(\chi)$ as follows:

\begin{theorem}\label{theorem:impact-kappa}
When $\CoeffPrate$ increases, the following claims are true
for each case under consideration:
\begin{enumerate}
\item
$\optEE(\chi)$ strictly increases.

\item
$\optSE(\chi)$ strictly decreases
if $\Prate(R)$ is strictly convex of $R\geq0$.
Moreover,
\begin{align}
\lim_{\CoeffPrate\rightarrow +\infty} \optSE(\chi) = 0; \;
\lim_{\CoeffPrate\rightarrow 0} \optSE(\chi) = \SE_1.  \label{eq:limit-optSE-kappa}
\end{align}
where $\SE_1$ satisfies that $\frac{\Nvar}{\PAEff}f(\SE_1) = \Pcst$.

\item
$\optSE(\chi)$ is fixed as $\SE_1$ if $\Prate(R) = R$.
\end{enumerate}
\end{theorem}
\begin{IEEEproof}
See the Appendix.
\end{IEEEproof}

Note that when $\CoeffPrate=0$, the power amplifier's power consumption
is the only source for the rate-dependent circuit power,
and $\optSE(\chi)=\SE_1$ in such a case.
Theorem \ref{theorem:impact-kappa} reveals important insight that
the behavior of $\optSE(\chi)$ when $\CoeffPrate$ increases
depends on the specific model that $\Prate(R)$ follows.
If $\Prate(R)=R$, $\optSE(\chi)$ keeps unchanged as $\SE_1$
when $\CoeffPrate$ increases.
However, $\optSE(\chi)$ decreases and approaches zero
if $\Prate(R)$ is strictly convex of $R$.
It is also interesting to see that $\optSE(\chi)$ when $\Prate(R)=R$
is always higher than that when $\Prate(R)$ is strictly convex of $R$.

%=================== impact-Nvar =====================
The impact of $\Nvar$ on $\optSE(\chi)$ and $\optEE(\chi)$
is shown as follows:

\begin{theorem}\label{theorem:impact-Nvar}
When $\Nvar$ increases, the following claims are true
for each case under consideration:
\begin{enumerate}
\item
$\optEE(\chi)$ strictly increases.

\item
$\optSE(\chi)$ strictly decreases. Moreover,
\begin{align}
\lim_{\Nvar\rightarrow +\infty} \optSE(\chi) = 0; \;
\lim_{\Nvar\rightarrow 0} \optSE(\chi) = \SE_2.  \label{eq:limi-optSE-Nvar}
\end{align}
where $\SE_2$ satisfies that $\CoeffPrate g(\SE_2) = \Pcst$.
\end{enumerate}
\end{theorem}
\begin{IEEEproof}
This theorem can be proven in a similar way as
Theorem \ref{theorem:impact-kappa}, thus the proof
is omitted here.
\end{IEEEproof}

Note that when $\Nvar=0$, $\CoeffPrate\Prate(R)$
is the only source for the rate-dependent circuit power,
and $\optSE(\chi)=\SE_2$.
Theorem \ref{theorem:impact-Nvar} indicates
when $\Nvar$ increases, $\optSE(\chi)$ decreases and
approaches zero.

%=================== impact-Pcst =====================
The impact of $\Pcst$ on $\optSE(\chi)$ and $\optEE(\chi)$
is shown as follows:

\begin{theorem}\label{theorem:impact-Pcst}
When $\Pcst$ increases, the following claims are true
for each case under consideration:
\begin{enumerate}
\item
$\optEE(\chi)$ strictly increases.

\item
$\optSE(\chi)$ strictly increases. Moreover,
\begin{align}
\lim_{\Pcst\rightarrow +\infty} \optSE(\chi) = +\infty; \;
\lim_{\Pcst\rightarrow 0} \optSE(\chi) = 0.  \label{eq:limit-optSE-Pcst}
\end{align}
\end{enumerate}
\end{theorem}
\begin{IEEEproof}
See the Appendix.
\end{IEEEproof}

For Case 1, we can also show the impact of $G$ on
$\optEE(\chi)$ and $\optSE(\chi)$ as follows:

\begin{theorem}
For Case 1, the following claims are true when $G$ increases:
\begin{enumerate}
\item
$\optEE(\chi)$ strictly decreases.

\item
$\optSE(\chi)$ strictly increases. Moreover,
\begin{align}
\lim_{G\rightarrow +\infty} \optSE(\chi) = \SE_2; \;
\lim_{G\rightarrow 0} \optSE(\chi) = 0.  \label{eq:limit-optSE-Pcst}
\end{align}
\end{enumerate}
\end{theorem}
\begin{IEEEproof}
This theorem can be proven in a similar way as Theorem \ref{theorem:impact-kappa},
thus the proof is omitted here.
\end{IEEEproof}

The above three theorems reveal important insight
about the impact of system parameters on
$\optSE(\chi)$ and $\optEE(\chi)$ for three cases under consideration.
For each case, the optimum EE always increases
when any one of $\Nvar$, $\CoeffPrate$ and $\Pcst$ increases.
For Case 1, the optimum EE degrades when $G$ decreases.
However, the behavior of the optimum SE depends on
the specific parameter that changes:
\begin{enumerate}
\item
If $\CoeffPrate$ increases which leads to the increase of
the rate-dependent circuit power consumption,
the link should keep rate unchanged if $\Prate(R)=R$,
to operate with the optimum EE.
This observation was also mentioned in \cite{Xiong11}
for the static channel case with CSIT.
However, if $\Prate(R)$ is strictly convex of $R\geq0$,
the link should slower its transmission\footnote{
In \cite{Xiong11} the linear rate-dependent circuit power
model is considered. At the end of Section III.A in \cite{Xiong11},
it was said that the linear rate-dependent circuit power model (i.e., $\kappa R$)
used there can be generalized to a convex model (i.e., $\kappa\phi(R)$)
as in our work for the static channel with CSIT.
However, it was claimed there that after the generalization the optimum SE
is still independent of $\kappa$

Here we show a different finding that
if $\phi(R)$ is strictly convex of $R$,
the optimum SE decreases as $\kappa$ increases
for all the three cases of the flat-fading channels.
The theoretical proof as well as simulation results (see Section VI)
are given to corroborate the above new finding.}.

\item
If $\Nvar$ increases (or $G$ decreases for Case 1)
which leads to the increase of
the power amplifier's power consumption,
the optimum SE has to be decreased, i.e., the link should
slower its transmission to operate with the optimum EE.

\item
If the rate-independent circuit power $\Pcst$ increases,
the optimum SE has to be increased, i.e., the link should transmit
at a higher rate to operate with the optimum EE.
This observation was also mentioned in \cite{Xiong11}
when $\Prate(R)=R$ for the static channel case with CSIT.
\end{enumerate}

It is also interesting to compare $\optEE(\chi)$ of the three channel cases.
It can readily be shown that
\begin{enumerate}
\item
$\optEE(\chi)$ for Case 1 is not higher (i.e., better or same)
than that for Case 2
if the average value of the random channel power gain (i.e., $\Avg{G}{G}$)
for Case 2 is the same as the fixed channel power gain
(i.e., $G$) for Case 1.
This is because when using the same transmission power $p$,
the SE for Case 2 is not higher than that for Case 1
due to the Jensen's inequality
(i.e., $\Avg{G}{\log_2(1 + G\frac{p}{\Nvar})}\leq
\log_2(1 + \Avg{G}{G}\frac{p}{\Nvar})$).

\item
$\optEE(\chi)$ for Case 3 is not higher than that for Case 2
when the random channel power gain for both cases
has the same distribution.
The reason is that for any given $\SE$, the total power
for Case 3 is not higher than that for Case 2
since for Case 3 the transmitter has the flexibility
to adapt the transmit power according to the available CSI.
\end{enumerate}

We will compare the $\optEE(\chi)$ for Case 1
with that for Case 3 by simulation results
as will be shown in Section \ref{sec:simulation}.

\section{Algorithm design}

When the circuit power is a constant, i.e., $\CoeffPrate=0$,
algorithms have been proposed in \cite{Ish12} to
find $\optSE(\chi)$ and $\optEE(\chi)$ for Cases 1 and 3.
When $\Prate(R)=R$, the algorithm proposed in \cite{Xiong11}
can be used to find $\optSE(\chi)$ and $\optEE(\chi)$ for Case 1.
However, they are not applicable when $\Prate(R)$ is
a general convex function of $R$ as considered in this paper.

For the three channel cases under consideration, we propose an algorithm,
which is summarized as Algorithm \ref{alg:find-optSE},
to find $\optSE(\chi)$ and $\optEE(\chi)$
based on the bisection method according to Lemma \ref{lemma:judge-SE}.
The key to this algorithm is to evaluate
$\Gamma_\chi(\SE) = \CoeffPrate g(\BW\SE) + \frac{\Nvar}{\PAEff}f(\SE) - \Pcst$
corresponding to any given $\SE$ for each case.
To this end, $g(\BW\SE) = \BW\SE\Prate'(\BW\SE) - \Prate(\BW\SE)$
can be computed according to the expression of $\Prate(R)$.
Moreover, the following procedures can be taken
to compute $f(\SE)$ for each case:
\begin{itemize}
\item
Case 1: it has been shown that
$\PowToNvar(\SE) = \frac{1}{G}(2^\SE - 1)$ and
$\PowToNvar'(\SE) = \frac{\ln2}{G}2^\SE$,
which means that
\begin{align}
f(\SE) = \frac{1}{G}[((\ln2)\SE-1)2^\SE + 1].  \label{eq:f-Case1}
\end{align}

\item
Case 2: note that $\PowToNvar(\SE)$ can be numerically evaluated
as the $p^\star$ satisfying
\begin{align}
\Avg{G}{\log_2(1 + Gp^\star)}=\SE
\end{align}
with the bisection method. It can readily be shown that
\begin{align}
\PowToNvar'(\SE) = \frac{\ln2}{\Avg{G}{\frac{G}{1+Gp^\star}}}.
\end{align}

This means that
\begin{align}
f(\SE) = \SE\frac{\ln2}{\Avg{G}{\frac{G}{1+Gp^\star}}} - p^\star.
\label{eq:f-Case2}
\end{align}

\item
Case 3: note that problem \eqref{eq:PowToNvar-SE-case3} is a convex
optimization problem. By introducing $\mu$ as the Lagrange
multiplier for the constraint, it can readily be seen that
the optimum $p(G)$ is
\begin{align}
p(G) = \left[\frac{\mu^\star}{\ln2} - \frac{1}{G}\right]^+,
\end{align}
where $\mu^\star$ is the nonnegative value satisfying
\begin{align}
\Avg{G}{\log_2(1 + Gp(G))}=\SE.  \label{eq:condition-Case3}
\end{align}

Moreover, $\mu^\star$ is equal to the increasing rate
of the optimum objective value for problem \eqref{eq:PowToNvar-SE-case3}
with respect to $\SE$ according to the sensitivity analysis
in convex optimization theory
(see pages 249-253 of \cite{Convex-opt} for more details.)
This means that $\PowToNvar'(\SE) = \mu^\star$ holds.
Note that we have already used this {\it sensitivity-analysis} based 
optimization technique in previous works, 
e.g. \cite{Wang13ICCC,Wang11JSAC,Wang13TSP,Wang13CL}, 
to design resource allocation algorithms in a very efficient and effective way.
As a result, $f(\SE)$ can be evaluated according to
\begin{align}
f(\SE) = \SE\mu^\star - \Avg{G}{p(G)}.   \label{eq:f-Case3}
\end{align}
\end{itemize}

\begin{algorithm}
\caption{The algorithm to compute $\optSE$ and $\optEE$ for each case.} \label{alg:find-optSE}
\begin{algorithmic}[1]
\STATE $\minSE = 0$; $\maxSE = 1$;
\STATE evaluate $\Gamma_\chi(\maxSE)$ by using \eqref{eq:f-Case1},
               \eqref{eq:f-Case2} and \eqref{eq:f-Case3} for Cases 1, 2 and 3, respectively;
\WHILE{$\Gamma_\chi(\maxSE)<0$}
    \STATE $\maxSE = 2*\maxSE$
\ENDWHILE

\WHILE{$\maxSE - \minSE>\delta$}
       \STATE  $\SE = 0.5(\maxSE + \minSE)$;
       \STATE  evaluate $\Gamma_\chi(\SE)$ by using \eqref{eq:f-Case1},
               \eqref{eq:f-Case2} and \eqref{eq:f-Case3} for Cases 1, 2 and 3, respectively;
       \IF{$\Gamma_\chi(\SE)=0$}
          \STATE  go to line 15;
       \ELSIF{$\Gamma_\chi(\SE)>0$}
          \STATE  $\maxSE = \SE$;
       \ELSE
          \STATE  $\minSE = \SE$;
       \ENDIF
\ENDWHILE
\STATE  output $\optSE(\chi) = \SE$ and
        $\optEE(\chi) = \EE(\optSE(\chi))$.
\end{algorithmic}
\end{algorithm}

In Algorithm \ref{alg:find-optSE},
$\delta>0$ is a prescribed small value to terminate the iteration.
It can readily be shown that the worst-case complexity of
Algorithm \ref{alg:find-optSE} is
$O\big(\log_2(\frac{1}{\delta})\big)$.

%The idea is to solve for the optimum $\optSE$
%and use
%Specifically, the optimization algorithm is designed by
%treating the power allocation as optimization variables,
%and it relies on first finding
%the $\lambda^\star$ satisfying $F(\lambda^\star) = 0$, where
%\begin{align}
%F(\lambda) = \max_{p}\hspace{0.2cm} \BW\SE(p) - \lambda[\CoeffPrate\Prate(\BW\SE(p))
%                                   + \frac{p}{\PAEff} + \Pcst],
%\end{align}
%and then the optimum $p$ for the right-hand side problem
%is the optimum $p$.
%However, in the general case $\Prate(\BW\SE(p)$ is
%neither convex or concave of $p$,
%which means the above right-hand side problem is not convex problem.

\section{Simulation results}\label{sec:simulation}

In order to corroborate the insight obtained
from the theoretical analysis,
we have implemented Algorithm \ref{alg:find-optSE}
to compute $\optSE(\chi)$ and $\optEE(\chi)$ for each case,
and carried out simulations for a typical flat-fading
communication link whose parameters take practical values
very close to those used in \cite{Cui05}.
Specifically, $\BW = 10$ KHz and $\Nvar = \BW N_0 N_f$
are used where $N_0 = -170$ dBm/Hz is the noise
power spectral density and $N_f$ is the noise figure.
The efficiency of the power amplifier efficiency
is set as $\PAEff = 0.4$.
When using Algorithm \ref{alg:find-optSE} in the simulations,
we set $\delta=10^{-8}$ and convergence of the algorithm
is always observed.

The average channel power gain is evaluated according to
\begin{align}
\overline{G} = G_0 d^{-3.5}
\end{align}
where $d$ is the transmitter-receiver distance,
and $G_0 = -70$ dB is chosen as in \cite{Cui05}.
$|h|$ is generated as follows:
\begin{enumerate}
\item
For Case 1 (static channel with CSIT),
$|h|$ is fixed as $\sqrt{\overline{G}}$.
\item
For Case 2 (FV channel with CDIT) and Case 3 (FV channel with CSIT):
$|h|$ is assumed to follow Nakagami distribution with parameter $m$ and $\Avg{h}{|h|^2} = \overline{G}$.
We choose this distribution because the parameter $m$
can easily be adjusted to reflect the severity of fast fading:
the larger $m$ indicates less severe fading, i.e., the pdf of $G=|h|^2$
is more compactly concentrated around its average value
$\overline{G}$. Specifically, the Nakagami distribution when $m=1$
is simply the Rayleigh distribution.
\end{enumerate}

\subsection{Illustration of the impact
of $\CoeffPrate$ on $\optSE(\chi)$ and $\optEE(\chi)$}

\begin{figure}
  \centering
     \subfigure[when $\Prate(R)=R$]{
         \includegraphics[width=3.5in]{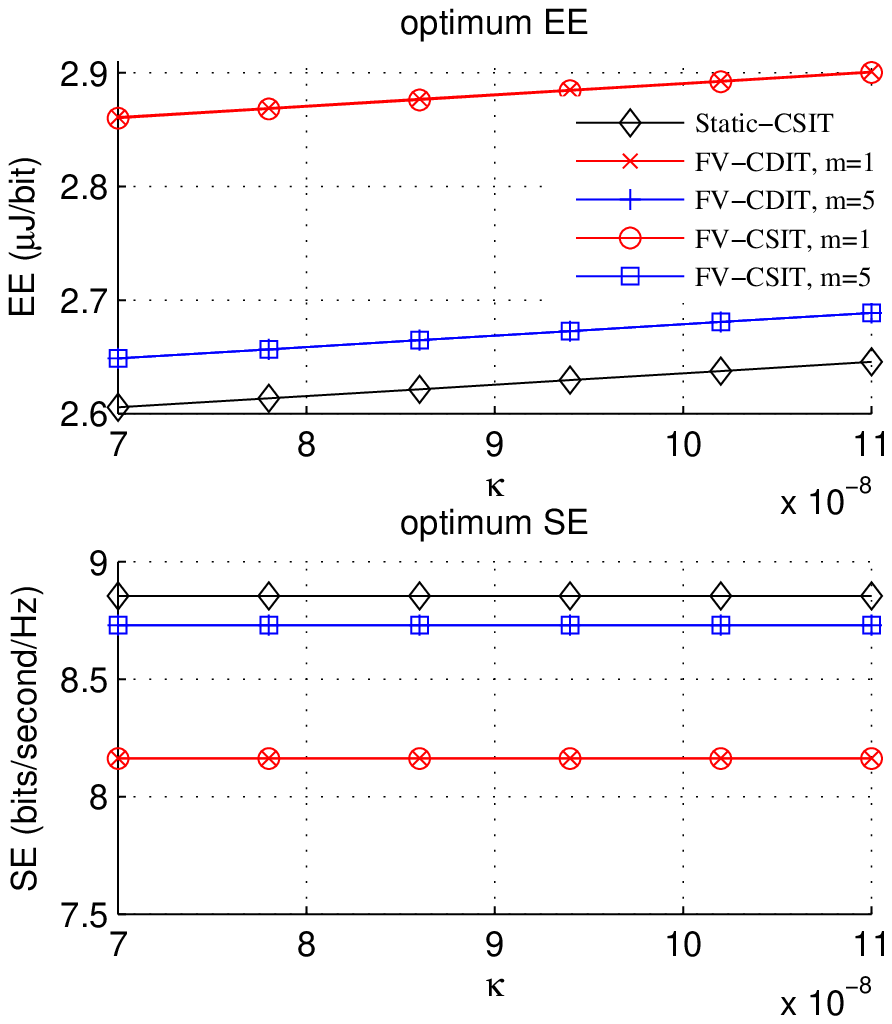}}
     \subfigure[when $\Prate(R)=R^{1.3}$]{
         \includegraphics[width=3.5in]{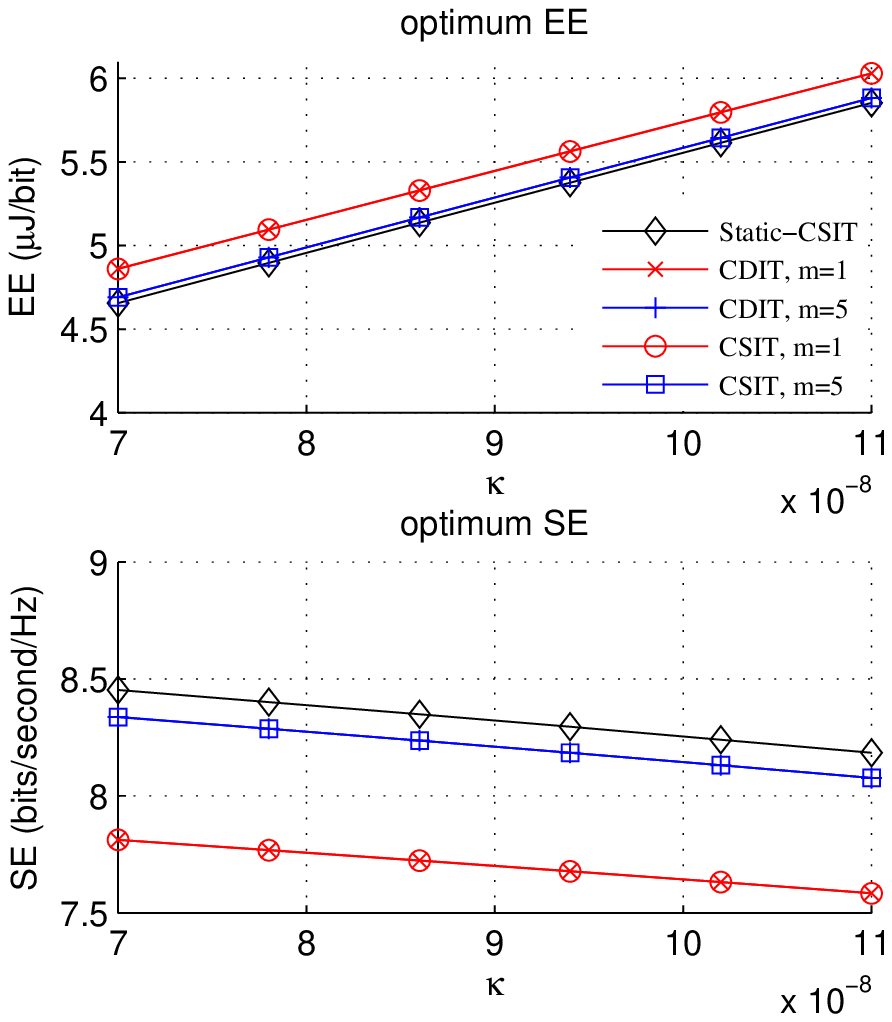}}
   \caption{The computed $\optEE(\chi)$ and $\optSE(\chi)$
           when $\CoeffPrate$ increases from $7\times10^{-8}$ to $1\times10^{-7}$,
           $d=10$ meter, $N_f=10$ dB, $\Pcst=188$ mW.}
  \label{fig:impact-kappa}
\end{figure}

To corroborate the insight about the impact of $\CoeffPrate$
on $\optEE(\chi)$ and $\optSE(\chi)$,
we assume $d=10$ meter, $N_f=10$ dB and $\Pcst=188$ mW which
are the same as that used in \cite{Cui05}.
$\optEE(\chi)$ and $\optSE(\chi)$ for each channel case
have been computed when $\Prate(R)=R$ and $\CoeffPrate$ increases from\footnote{
This means that the rate-dependent circuit energy
consumption per bit increases from $70$ to $100$ nJ,
which agrees with those reported in \cite{ranpara1999low,zhang2010efficient}.}
$7\times10^{-8}$ to $1\times10^{-7}$.
We have also computed $\optEE(\chi)$ and $\optSE(\chi)$
for each channel case with the same parameters except for $\Prate(R)=R^{1.3}$.
The results are shown in Figure \ref{fig:impact-kappa}.

It can be seen that as $\CoeffPrate$ increases,
$\optEE(\chi)$ increases regardless of the form that $\Prate(R)$ takes.
We have also evaluated by the numerical method that
the $\SE_1$ satisfying $\frac{\Nvar}{\PAEff}f(\SE_1)=\Pcst$
is equal to $8.85$, $8.73$ and $8.16$
for Cases 1, 2 and 3, respectively.
It is shown in Fig. \ref{fig:impact-kappa} that
the $\optSE(\chi)$ for the three channel cases when $\Prate(R)=R$
keeps fixed at those values when $\CoeffPrate$ increases.
Moreover, $\optSE(\chi)$ for each channel case when $\Prate(R)=R^{1.3}$
(i.e., $\Prate(R)$ is strictly convex of $R$)
decreases and is always smaller than the $\optSE(\chi)$
when $\Prate(R)=R$.
These observations corroborate the insight
obtained in Section \ref{sec:impact}.

When $m$ is fixed, $\optEE(\chi)$ for Case 2 (FV channel with CDIT)
is always higher than that for Case 1 (static channel with CSIT),
which is in agreement with the theoretical analysis in Section \ref{sec:impact}.
When $m$ increases,
$\optEE(\chi)$ for Case 2 approaches that for Case 1.
The is because when $m$ takes a large value,
the pdf of $G$ becomes more compactly concentrated
around the average value $\overline{G}$, hence
$\Avg{G}{\log_2(1 + G p)} \approx \log_2(1 + \overline{G}p)$
holds, meaning that $\PowToNvar(\SE)$ for Case 2
is very close to that for Case 1.
These observation indicates that the less severe fading
leads to improved EE performance for Case 2.

It can also be seen that
$\optEE(\chi)$ for Case 3 (FV channel with CSIT) is very close
to that for Case 2, which is explained as follows.
Note that $\optSE(\chi)$ for Case 2 and Case 3 are
very close and relatively high.
For Case 3, the corresponding $\mu^\star$
has to take a high value in order to satisfy \eqref{eq:condition-Case3}
when $\SE=\optSE(\chi)$.
In such a case, the optimum $p(G)$ is approximately
a constant, thus $\PowToNvar(\optSE(\chi))$ for Case 3
is very close to $\PowToNvar(\optSE(\chi))$ for Case 2.
Therefore, the $\optEE(\chi)$ for the two cases are approximately equal.

\subsection{Illustration of the impact
of $\Nvar$ on $\optSE(\chi)$ and $\optEE(\chi)$}

\begin{figure}
  \centering
     \includegraphics[width=3.5in]{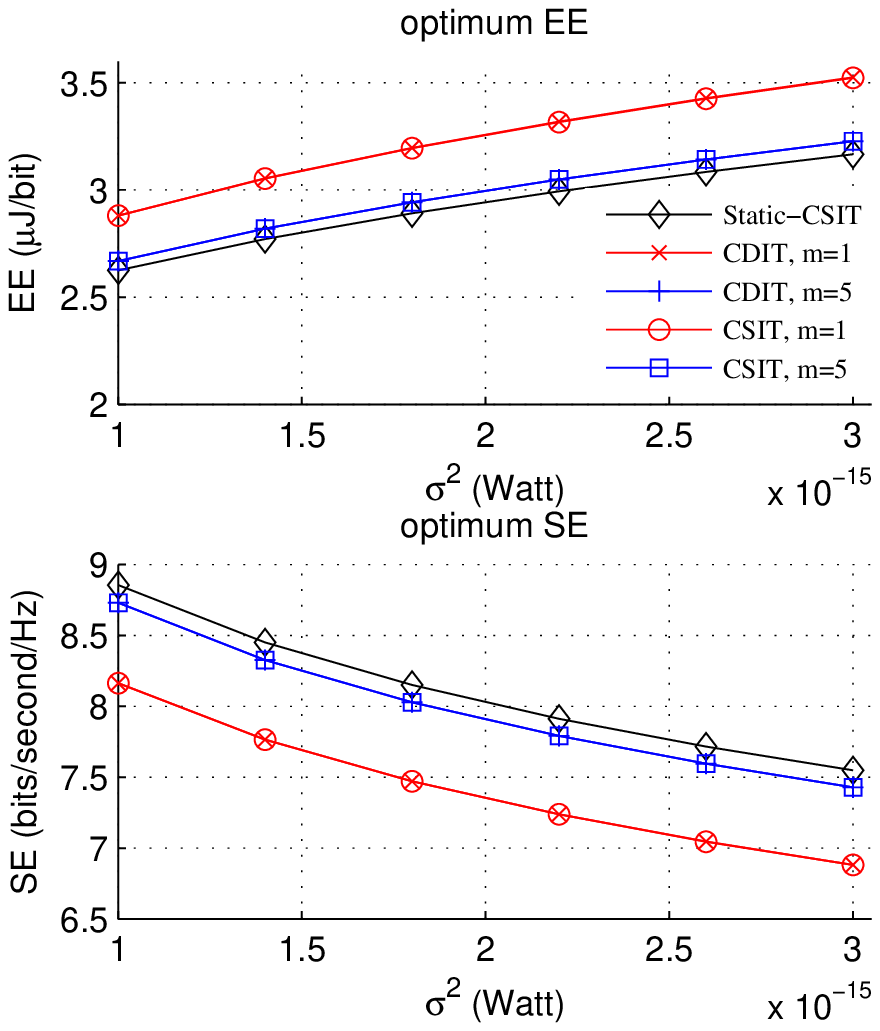}
  \caption{The computed $\optEE(\chi)$ and $\optSE(\chi)$
           when $N_f$ increases from $10$ dB to $30$ dB,
           $d=10$ meter, $\Pcst=188$ mW, $\CoeffPrate=9\times10^8$, $\Prate(R)=R$.}
  \label{fig:impact-Nvar}
\end{figure}

To corroborate the insight about the impact of $\Nvar$
on $\optEE(\chi)$ and $\optSE(\chi)$,
we assume $d=10$ meter, $\Pcst=188$ mW,
$\CoeffPrate=9\times10^8$ and $\Prate(R)=R$.
$\optEE(\chi)$ and $\optSE(\chi)$ for each case
have been computed when $N_f$ increases from $10$ to $30$ dB.
The results are shown in Figure \ref{fig:impact-Nvar}.
It can be seen that as $\Nvar$ increases due to the increase of $N_f$,
$\optEE(\chi)$ increases while $\optSE(\chi)$ decreases,
which corroborate the insight obtained in Section \ref{sec:impact}.
Moreover, similar points can be observed as said earlier
when comparing $\optEE(\chi)$ of the three channel cases.

\subsection{Illustration of the impact
of $\Pcst$ on $\optSE(\chi)$ and $\optEE(\chi)$}

\begin{figure}
  \centering
     \includegraphics[width=3.5in]{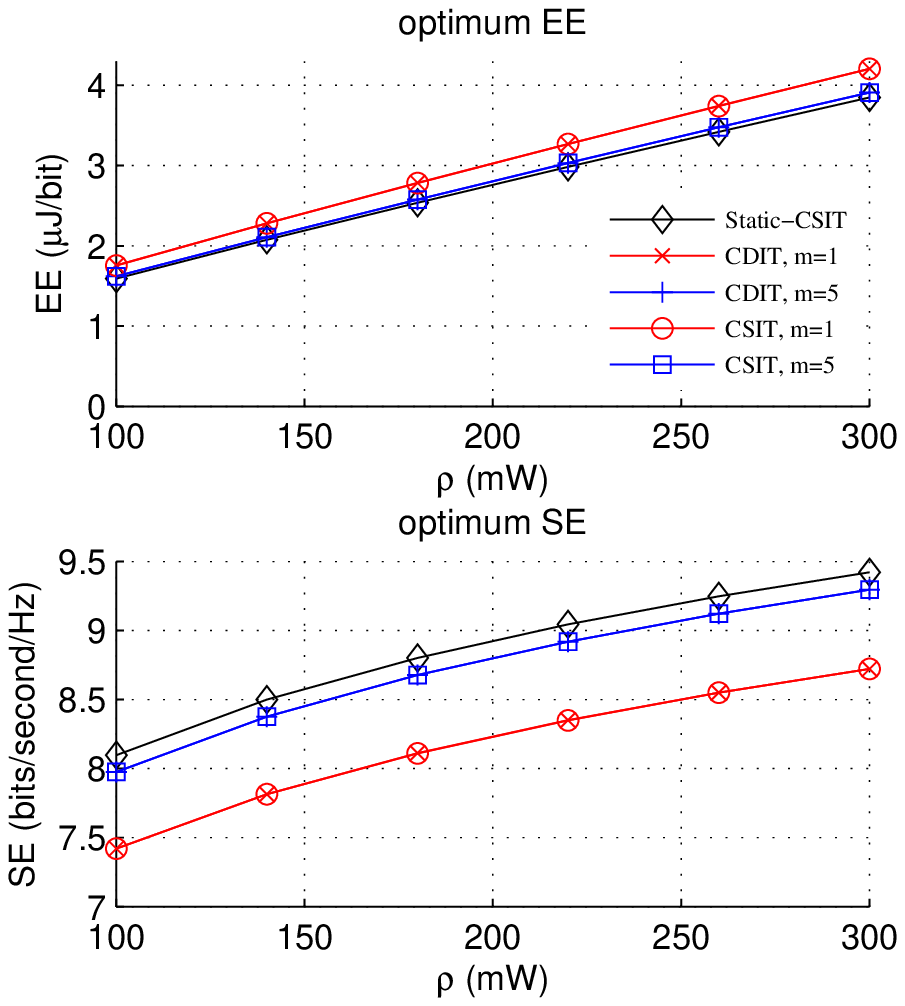}
  \caption{The computed $\optEE(\chi)$ and $\optSE(\chi)$
           when $\Pcst$ increases from $100$ mW to $300$ mW,
           $d=10$ meter, $N_f$ = 10 dB, $\CoeffPrate=9\times10^8$, $\Prate(R)=R$.}
  \label{fig:impact-Pcst}
\end{figure}

To corroborate the insight about the impact of $\Pcst$
on $\optEE(\chi)$ and $\optSE(\chi)$,
we assume $d=10$ meter, $N_f=10$ dB,
$\CoeffPrate=9\times10^8$ and $\Prate(R)=R$.
$\optEE(\chi)$ and $\optSE(\chi)$ for each case
have been computed when $\Pcst$ increases from $100$ to $300$ mW.
The results are shown in Figure \ref{fig:impact-Pcst}.
It can be seen that as $\Pcst$ increases,
both $\optEE(\chi)$ and $\optSE(\chi)$ increase,
which corroborate the insight obtained in Section \ref{sec:impact}.
Moreover, similar points can be observed as said earlier
when comparing $\optEE(\chi)$ of the three channel cases.

\subsection{Illustration of the impact of
$G$ on $\optSE(\chi)$ and $\optEE(\chi)$}

\begin{figure}
  \centering
  \subfigure[when $d$ increases from $10$ to $30$ meter]{   \includegraphics[width=3.5in]{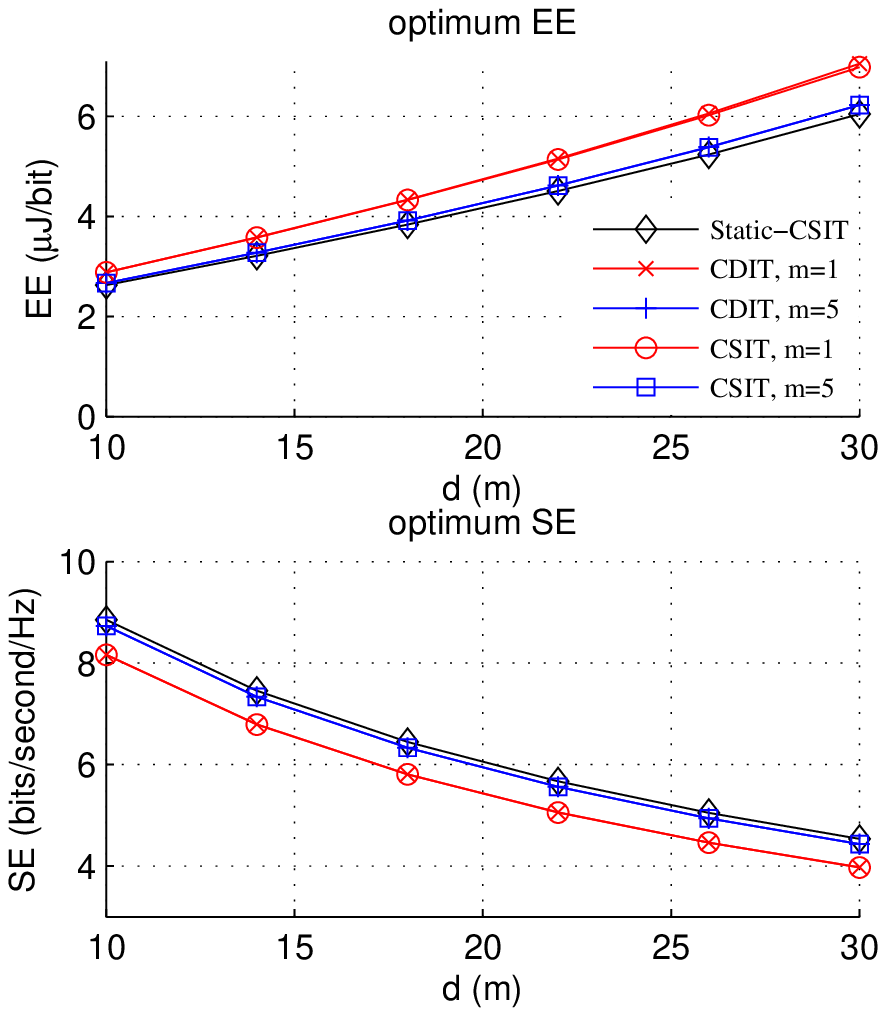}}
  \subfigure[when $d$ increases from $150$ to $160$ meter]{
  \includegraphics[width=3.5in]{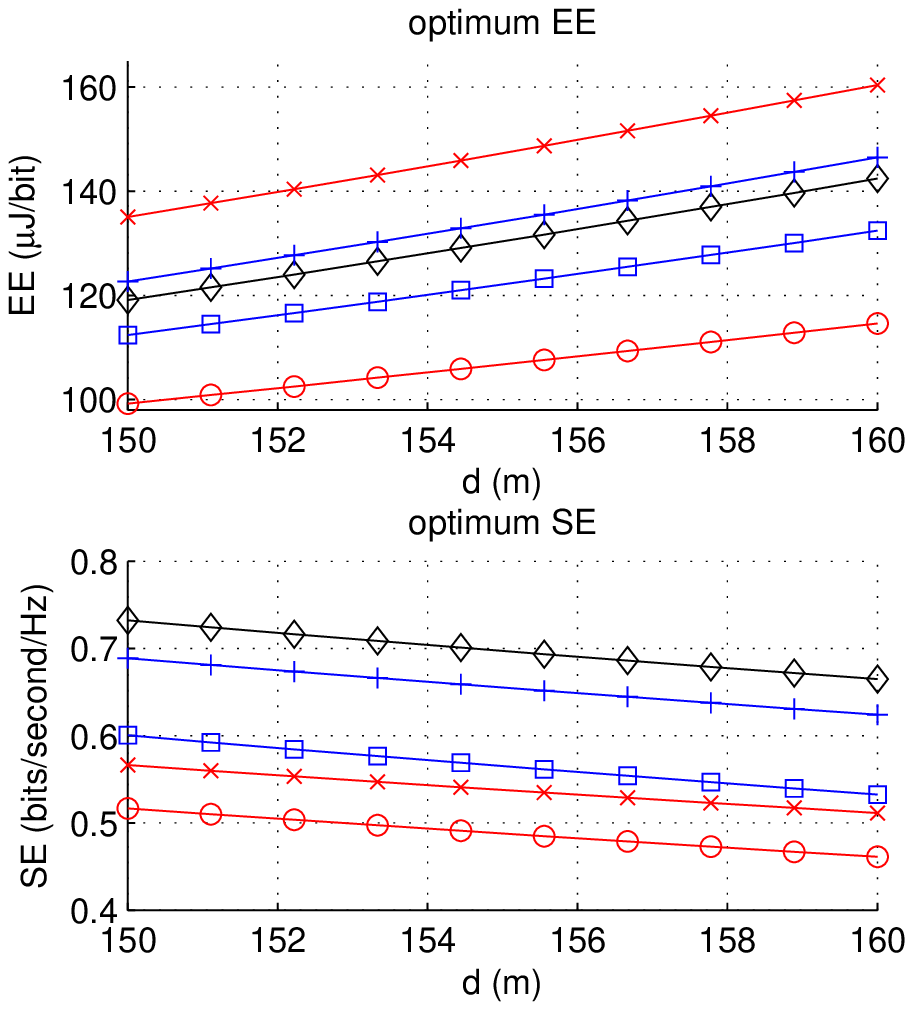}}
  \caption{The computed $\optEE(\chi)$ and $\optSE(\chi)$
           when $\Pcst=188$ mW, $N_f$ = 10 dB, $\CoeffPrate=9\times10^8$
           and $\Prate(R)=R$.}
  \label{fig:impact-dist}
\end{figure}

To corroborate the insight about the impact of $G$
on $\optEE(\chi)$ and $\optSE(\chi)$ for Case 1,
we assume $N_f=10$ dB, $\Pcst=188$ mW,
$\CoeffPrate=9\times10^8$ and $\Prate(R)=R$.
$\optEE(\chi)$ and $\optSE(\chi)$ for each channel case
have been computed when $d$ increases from $10$ to $30$ meter.
We have also computed $\optEE(\chi)$ and $\optSE(\chi)$
for each channel case when $d$ increases from $150$ m to $160$ meter.
The results are shown in Figure \ref{fig:impact-Nvar}.
It can be seen that as $G$ decreases due to the increase of $d$,
$\optEE(\chi)$ increases while $\optSE(\chi)$ decreases,
which corroborate the insight obtained in Section \ref{sec:impact}.
When $d$ is between $10$ and $30$ m, points similar as said earlier
can be observed when comparing $\optEE(\chi)$ of the three channel cases.

However, when $d$ is between $150$ and $160$ meter,
different and interesting points can be observed.
In such a case, $\optSE(\chi)$ is relatively small for
all three channel cases.
$\optSE(\chi)$ for Case 2 is always higher than that for Case 1
and as $m$ increases, $\optSE(\chi)$ for Case 2 approaches that for Case 1,
meaning that the less severe fading leads to improved
EE performance for Case 2 as observed earlier.

However, $\optEE(\chi)$ for Case 3 is
smaller than $\optEE(\chi)$ for Case 1.
The reason is that since $\optSE(\chi)$ is small,
the corresponding average power is also small,
and thus the optimum power allocation $p(G)$ for Case 3
is to allocate power and activate transmission only when $G$ is high.
In such an opportunistic way, the optimum power allocation for Case 3
can exploit the channel fading in a very efficient way,
in the sense that only the channel states with high channel
power gain (which is very likely much higher than $\overline{G}$)
are used for data transmission.
As $m$ increases,
$\optSE(\chi)$ for Case 3 approaches that for Case 1,
because the pdf of $G$ becomes more compactly
concentrated around $\overline{G}$ so that
$G$ appears with a higher probability at values
very close to $\overline{G}$.
This means that the less severe fading
leads to degraded EE performance for Case 3
when $\optSE(\chi)$ is small as shown here.

\section{Conclusion}

We have investigated the optimum EE and corresponding
SE for a communication link over a flat-fading channel.
Three cases for the flat-fading channel
are considered, namely static channel with CSIT,
FV channel with CDIT and FV channel with CSIT.
The link's circuit power is modeled as
the sum of a constant and a rate-dependent part
as an increasing and convex function of the transmission rate.
For all three cases,
the tradeoff between the EE and the spectral efficiency (SE)
has been studied, and the impact of system parameters
on the optimum EE and corresponding SE has been investigated
to obtain insight.
A polynomial-complexity algorithm has been developed with the
bisection method to find the optimum SE.
The insight has been corroborated and the optimum EE for
the three cases has been compared by simulation results.
{The insight and algorithm presented in this paper can be applied
to guide the practical design of communication links
over flat-fading channels for improved EE performance
as illustrated by simulation results.
In future, we will extend the study to multiuser networks
with frequency-selective channels as studied in \cite{zhu2009chunk1,zhu2012chunk2}.}

\section*{Acknowledgement}

The authors would like to thank Prof. Jiangzhou Wang
as well as the anonymous reviewers for their valuable
comments to improve the quality of this paper.

\appendices
\appendix

\subsection{Proof of Lemma \ref{lemma:PowToNvar-SE-convex}} \label{proof:lemma1}

For Case 1, it can readily be seen from \eqref{eq:PowToNvar-SE-case1}
that $\PowToNvar(\SE)$ is strictly increasing and strictly
convex of $\SE\geq0$.

The claims for Case 2 and Case 3 can be proven in a similar way,
hence we only prove the one for Case 3 as follows.
Obviously $\PowToNvar(\SE)$ is strictly increasing of $\SE\geq0$.
Proving its strictly convexity is equivalent to show that
$\forall\;\SE_1,\SE_2\geq0$, $\forall\;\alpha\in(0,1)$,
$\PowToNvar(\alpha\SE_1 + (1-\alpha)\SE_2)\leq
\alpha\PowToNvar(\SE_1) + (1-\alpha)\PowToNvar(\SE_2)$.
To this end, suppose $\Pstrat_1 = \{p_1(G)|\forall\;G\geq0\}$,
$\Pstrat_2=\{p_2(G)|\forall\;G\geq0\}$ and
$\Pstrat_3=\{p_3(G)|\forall\;G\geq0\}$ represent
the optimum $\Pstrat$ for problem \eqref{eq:PowToNvar-SE-case3}
when $\SE=\SE_1$, $\SE=\SE_2$ and $\SE=\alpha\SE_1+(1-\alpha)\SE_2$, respectively.
This means that $\PowToNvar(\SE_i)=\Avg{G}{p_i(G)}$
where $i\in\{1,2,3\}$.

Define $f(\Pstrat)=\Avg{G}{\log_2(1 + Gp(G))}$
where $\Pstrat=\{p(G)|\forall\;G\geq0\}$.
It can readily be shown that $f(\Pstrat)$
is strictly concave of $\Pstrat\in\Pspace$.
Obviously, the constraint of problem \eqref{eq:PowToNvar-SE-case3}
must be saturated at the optimum solution, i.e.,
$f(\Pstrat_1)=\SE_1$, $f(\Pstrat_2)=\SE_2$
and $f(\Pstrat_3)=\alpha\SE_1+(1-\alpha)\SE_2$ must hold.
According to the strict concavity of $f(\Pstrat)$,
\begin{align}
f(\alpha\Pstrat_1 + (1-\alpha)\Pstrat_2)
&> \alpha f(\Pstrat_1) + (1-\alpha)f(\Pstrat_2)  \nonumber\\
&= \alpha\SE_1 + (1-\alpha)\SE_2                 \nonumber
\end{align}
follows. This means that $\alpha\Pstrat_1 + (1-\alpha)\Pstrat_2$
is a feasible but not optimum solution for problem \eqref{eq:PowToNvar-SE-case3}
with $\SE=\alpha\SE_1+(1-\alpha)\SE_2$.
On the other hand, $\Pstrat_3$ is the optimum solution
for the same problem. This means that
\begin{align}
\PowToNvar(\alpha\SE_1+(1-\alpha)\SE_2)
&= \Avg{G}{p_3(G)}                                         \nonumber\\
&< \Avg{G}{\alpha p_1(G) + (1-\alpha)p_2(G)}               \nonumber\\
&= \alpha\PowToNvar(\SE_1) + (1-\alpha)\PowToNvar(\SE_2),  \nonumber
\end{align}
which proves the claim.

\subsection{Proof of Lemma \ref{lemma:EE-SE-quasiconvex}} \label{proof:lemma2}

According to Proposition C9 in \cite{Quasi-convex},
$\EE(\SE)$ is strictly quasiconvex of $\SE\geq 0$ if
\begin{align}
\Pi(\gamma) = \{\SE\geq0|\EE(\SE) \leq \gamma\}   \nonumber
\end{align}
is a strictly convex set for any real value $\gamma$.

Note that $\forall\;\SE\geq0$, $\EE(\SE)>0$,
which means that $\Pi(\gamma)$ is empty if $\gamma\leq0$,
hence $\Pi(\gamma)$ is strictly convex since
no point lies on the contour of $\Pi(\gamma)$.
We now prove the strict convexity of $\Pi(\gamma)$ when $\gamma>0$.
In such a case,
\begin{align}
\Pi(\gamma) = \{\SE\geq0|f(\gamma,\SE)=P(\SE)-\gamma\BW\SE \leq 0\}.  \nonumber
\end{align}

Suppose $\underline{\SE}$ and $\overline{\SE}$
are any two points on the contour of $\Pi(\gamma)$.
Obviously $\underline{\SE}>0$ and $\overline{\SE}>0$
since $0\notin\Pi(\gamma)$ due to the fact that $f(\gamma,0)>0$.
$\forall\;\SE\in(\underline{\SE},\overline{\SE})$,
\begin{align}
f(\gamma,\SE)< \max\{f(\gamma,\underline{\SE}),f(\gamma,\overline{\SE})\} \leq 0  \nonumber
\end{align}
follows from the strict convexity of $f(\gamma,\SE)$ with respect to $\SE$,
meaning that any $\SE$ between any two points on the contour of $\Pi(\gamma)$
must lie in the interior of $\Pi(\gamma)$.
Thus, $\Pi(\gamma)$ is a strictly convex set when $\gamma>0$.
Therefore, $\EE(\SE)$ is strictly quasiconvex of $\SE$.

\subsection{Proof of Theorem \ref{theorem:EE-SE-tradeoff}}

To prove the first claim, suppose there exist $\SE'$ and $\SE''$
satisfying $\SE'<\SE''$ and $\EE(\SE') = \EE(\SE'') = \optEE$.
From \eqref{eq:quasiconvex-ineq}, $\forall\;\SE\in(\SE',\SE'')$,
$\optEE=\max\{\EE(\SE'),\EE(\SE'')\}>\EE(\SE)$, leading to a contradiction
with $\optEE \leq \EE(\SE)$.
Therefore, there must exist a unique $\optSE$
satisfying $\EE(\optSE)=\optEE$.
Moreover, $\optSE$ must satisfy
\begin{align}
\forall\;\SE\geq 0, \EE'(\optSE)(\SE-\optSE)\geq0   \label{eq:stationary-cond}
\end{align}
according to Proposition 2.1.2 in \cite{Nonlinear-opt}.
As said earlier, $\optSE>0$ must hold, thus
$\EE'(\optSE)=0$ must hold
to satisfy the condition \eqref{eq:stationary-cond}.
From \eqref{eq:Der-EE}, \eqref{eq:optSE-cond} must hold.
This proves the first claim.

We now prove the second claim.
For any $\SE_1$ and $\SE_2$ satisfying $0< \SE_1 < \SE_2 \leq \optSE$,
$\EE(\SE_1) = \max\{\EE(\SE_1), \EE(\optSE)\} > \EE(\SE_2)$
follows from \eqref{eq:quasiconvex-ineq}.
This means that $\EE(\SE)$ is strictly decreasing with $\SE\in(0,\optSE)$,
Therefore, $\EE'(\SE)<0$ must hold $\forall\;\SE\in(0,\optSE)$.
From \eqref{eq:Der-EE}, \eqref{eq:leftSE-cond} must hold.
This proves the second claim.

The third claim is proven as follows.
For any $\SE_1$ and $\SE_2$ satisfying $\SE_1 > \SE_2 \geq \optSE$,
$\EE(\SE_1) = \max\{\EE(\SE_1), \EE(\optSE)\} > \EE(\SE_2)$
follows from \eqref{eq:quasiconvex-ineq}.
This means that $\EE(\SE)$ is strictly increasing with $\SE\in(\optSE,+\infty)$,
Therefore, $\EE'(\SE)>0$ must hold $\forall\;\SE\in(\optSE,+\infty)$.
From \eqref{eq:Der-EE}, \eqref{eq:rightSE-cond} must hold.
This proves the third claim.

\subsection{Proof of Lemma \ref{lemma:judge-SE}}

According to Theorem \ref{theorem:EE-SE-tradeoff},
\begin{align}
\SE \hspace{0.1cm}\left\{\begin{array}{ll}
                              < \optSE(\chi)
                              &{\rm iff}\hspace{0.2cm} Z_\chi(\SE) < 0 \\
                              = \optSE(\chi)
                              &{\rm iff}\hspace{0.2cm} Z_\chi(\SE) = 0\\
                              > \optSE(\chi)
                              &{\rm iff}\hspace{0.2cm} Z_\chi(\SE) > 0
                          \end{array} \right.
                          \label{eq:judge-SE-2}
\end{align}
must hold where
\begin{align}
Z_\chi(\SE) &= P'(\SE) - \frac{P(\SE)}{\SE}     \nonumber\\
            &= \frac{1}{\SE}
                \left[\CoeffPrate[\BW\SE\Prate'(\BW\SE)-\Prate(\BW\SE)]
                      + \frac{\Nvar}{\PAEff}[\SE\PowToNvar'(\SE)-\PowToNvar(\SE)]
                      - \Pcst \right]        \nonumber\\
       &= \frac{1}{\SE}\left[\CoeffPrate g(\BW\SE)
                             + \frac{\Nvar}{\PAEff}f(\SE)
                             - \Pcst\right] = \frac{1}{\SE}\Gamma_\chi(\SE) \nonumber
\end{align}

From the above equation, it can readily be seen
\eqref{eq:judge-SE-2} is equivalent to \eqref{eq:judge-SE},
which proves the first claim.

We now prove the claim about $g(R)$.
Obviously $g(0)=0$ holds.
It can readily be verified that
$g(R)=0$, $\forall\;R\geq0$ if $\Prate(R) = R$.
If $\Prate(\SE)$ is strictly convex of $R\geq0$,
\begin{align}
g'(R) = \Prate'(R) + R\Prate''(R) - \Prate'(R) = R\Prate''(R),   \label{eq:der-fSE}
\end{align}
meaning that $g(R)$ is strictly increasing of $R\geq0$
due to the strict convexity of $\Prate(R)$.
Hence the claim about $g(R)$ is proven.
In a similar way it can be proven that
$f(\SE)$ is strictly increasing of $\SE\geq0$.

\subsection{Proof of Theorem \ref{theorem:impact-kappa}}

Suppose $\CoeffPrate$ increases from
$\CoeffPrate_1$ to $\CoeffPrate_2$ (i.e., $\CoeffPrate_2>\CoeffPrate_1$)
while all other parameters are fixed.
Denote the $\chi$ when $\CoeffPrate=\CoeffPrate_1$ and $\CoeffPrate=\CoeffPrate_2$
as $\chi_1$ and $\chi_2$, respectively.

To prove the first claim, note that
\begin{align}
\optEE(\chi_2) &= \frac{\CoeffPrate_2 \Prate(\BW\optSE(\chi_2))
+ \frac{\Nvar}{\PAEff}\PowToNvar(\optSE(\chi_2))
+ \Pcst}{\BW\optSE(\chi_2)}  \nonumber\\
&> \frac{\CoeffPrate_1 \Prate(\BW\optSE(\chi_2))
+ \frac{\Nvar}{\PAEff}\PowToNvar(\optSE(\chi_2))
+ \Pcst}{\BW\optSE(\chi_2)}  \nonumber\\
&\geq \frac{\CoeffPrate_1 \Prate(\BW\optSE(\chi_1))
+ \frac{\Nvar}{\PAEff}\PowToNvar(\optSE(\chi_1))
+ \Pcst}{\BW\optSE(\chi_1)}  \nonumber\\
& = \optEE(\chi_1),
\end{align}
where the inequality in the second line is due to the fact that
$\CoeffPrate_2>\CoeffPrate_1$ and $\optSE(\chi_2)>0$ (meaning that $\Prate(\BW\optSE(\chi_2))>0$).
The inequality in the third line is due to the fact that
$\optSE(\chi_2)$ is a feasible SE while $\optSE(\chi_1)$
is the optimum SE minimizing the EE when $\CoeffPrate=\CoeffPrate_1$.
This proves the first claim.

We now prove the second claim. Note that when
$\Prate(R)$ is strictly convex of $R\geq0$,
$g(R)$ is strictly increasing of $R\geq0$,
meaning that $g(\BW\optSE(\chi_1))>0$. Therefore,
\begin{align}
\Gamma_{\chi_2}(\optSE(\chi_1))
&= \CoeffPrate_2 g(\BW\optSE(\chi_1)
   + \frac{\Nvar}{\PAEff} f(\optSE(\chi_1)) - \Pcst  \nonumber\\
&> \CoeffPrate_1 g(\BW\optSE(\chi_1)
   + \frac{\Nvar}{\PAEff} f(\optSE(\chi_1)) - \Pcst  \nonumber\\
&= \Gamma_{\chi_1}(\optSE(\chi_1)) = 0,              \label{eq:inequality-SE-kappa}
\end{align}
follows. According to Lemma \ref{lemma:judge-SE},
$\optSE(\chi_1) > \optSE(\chi_2)$ holds.
This proves the claim that $\optSE(\chi)$ strictly decreases
when $\CoeffPrate$ increases.

To prove \eqref{eq:limit-optSE-kappa},
note that $\optSE(\chi)$ satisfies that
$\CoeffPrate g(\BW\optSE(\chi)) + \frac{\Nvar}{\PAEff}f(\optSE(\chi)) = \Pcst$.
On the one hand,
\begin{align}
\lim_{\CoeffPrate\rightarrow+\infty}g(\BW\optSE(\chi))
= \lim_{\CoeffPrate\rightarrow+\infty}\frac{\Pcst - \frac{\Nvar}{\PAEff}f(\optSE(\chi))}{\CoeffPrate} = 0  \nonumber
\end{align}
holds, meaning that $\lim_{\CoeffPrate\rightarrow+\infty}\optSE(\chi) = 0$.
On the other hand,
\begin{align}
\lim_{\CoeffPrate\rightarrow0}\frac{\Nvar}{\PAEff}f(\optSE(\chi))
= \lim_{\CoeffPrate\rightarrow0}[\Pcst - \CoeffPrate g(\BW\optSE(\chi))]
= \Pcst \nonumber
\end{align}
holds, meaning that $\lim_{\CoeffPrate\rightarrow0}\optSE(\chi) = \SE_1$.
This proves the \eqref{eq:limit-optSE-kappa}.

To prove the last claim, note that when $\Prate(R)=R$,
$\frac{\Nvar}{\PAEff}f(\optSE(\chi))=\Pcst$ holds since $g(\BW\optSE(\chi))=0$.
This means that $\optSE(\chi)=\SE_1$ always holds.

\subsection{Proof of Theorem \ref{theorem:impact-Pcst}}

Suppose $\Pcst$ increases from $\Pcst_1$ to $\Pcst_2$
(i.e., $\Pcst_2>\Pcst_1$) while all other parameters are fixed.
Denote the $\chi$ when $\Pcst=\Pcst_1$ and $\Pcst=\Pcst_2$
as $\chi_1$ and $\chi_2$, respectively.
Note that
\begin{align}
\optEE(\chi_2) &= \frac{\CoeffPrate \Prate(\BW\optSE(\chi_2))
+ \frac{\Nvar}{\PAEff}\PowToNvar(\optSE(\chi_2))
+ \Pcst_2}{\BW\optSE(\chi_2)}  \nonumber\\
&> \frac{\CoeffPrate \Prate(\BW\optSE(\chi_2))
+ \frac{\Nvar}{\PAEff}\PowToNvar(\optSE(\chi_2))
+ \Pcst_1}{\BW\optSE(\chi_2)}  \nonumber\\
&\geq \frac{\CoeffPrate \Prate(\BW\optSE(\chi_1))
+ \frac{\Nvar}{\PAEff}\PowToNvar(\optSE(\chi_1))
+ \Pcst_1}{\BW\optSE(\chi_1)}  \nonumber\\
& = \optEE(\chi_1),
\end{align}
follows, where the inequality in the third line
is due to the fact that
$\optSE(\chi_2)$ is a feasible SE while $\optSE(\chi_1)$
is the optimum SE minimizing the EE when $\Pcst=\Pcst_1$.
This proves the first claim.

We now prove the second claim. Note that
\begin{align}
\Gamma_{\chi_2}(\optSE(\chi_1))
&= \CoeffPrate g(\BW\optSE(\chi_1)
   + \frac{\Nvar}{\PAEff} f(\optSE(\chi_1)) - \Pcst_2  \nonumber\\
&< \CoeffPrate g(\BW\optSE(\chi_1)
   + \frac{\Nvar_1}{\PAEff} f(\optSE(\chi_1)) - \Pcst_1  \nonumber\\
&= \Gamma_{\chi_1}(\optSE(\chi_1)) = 0,              \label{eq:inequality-SE-kappa}
\end{align}
follows. According to Lemma \ref{lemma:judge-SE},
$\optSE(\chi_1) < \optSE(\chi_2)$ holds.
This proves that $\optSE(\chi)$ strictly increases
when $\Pcst$ increases.

To prove \eqref{eq:limit-optSE-Pcst},
note that $\optSE(\chi)$ is equal to the $\SE$ satisfying
$\CoeffPrate g(\BW\SE) + \frac{\Nvar}{\PAEff}f(\SE) = \Pcst$.
According to Lemma \ref{eq:judge-SE},
$\CoeffPrate g(\BW\SE) + \frac{\Nvar}{\PAEff}f(\SE)$ is a strictly
increasing function of $\SE$.
Therefore, \eqref{eq:limit-optSE-Pcst} holds.


\ifCLASSOPTIONcaptionsoff
  \newpage
\fi



\bibliographystyle{IEEEtran}
\bibliography{EE-literature}%

%\vfill


\begin{IEEEbiography}[{\includegraphics[width=1in,height=1.5in,clip,keepaspectratio]
{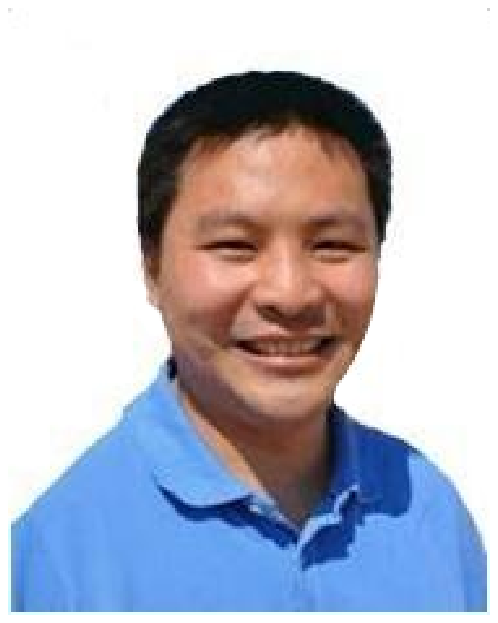}}]{Tao Wang}(SM'12)
received Ph.D from Universit{\'e} Catholique de Louvain (UCL), Belgium in 2012,
Doctor of Engineering and Bachelor ({\it summa cum laude}) degrees
from Zhejiang University, China, in 2006 and 2001, respectively.
He has been with Key Laboratory of Specialty Fiber Optics
and Optical Access Networks,
School of Communication \& Information Engineering,
Shanghai University, China as a Professor since Feb. 2013.

His current interest is in resource allocation
and adaptive modulation techniques for wireless communication
and signal processing systems.
He received Professor of Special Appointment (Eastern Scholar) Award
from Shanghai Municipal Education Commission,
as well as Best Paper Award in
2013 International Conference on Wireless Communications and Signal Processing.
He is an associate editor for {\it EURASIP Journal on Wireless Communications
and Networking} and {\it Signal Processing: An International Journal (SPIJ)},
as well as an editorial board member of {\it Recent Patents on Telecommunications}.
He served as a session chair in 2012 IEEE International Conference on Communications,
2013 IEEE \& CIC International Conference on Communications in China,
and 2013 International Conference on Wireless Communications and Signal Processing.
He was also a TPC member of International Congress on Image and Signal Processing
in 2012 and 2010. He is an IEEE Senior Member.
\end{IEEEbiography}

\begin{IEEEbiography}[{\includegraphics[width=1in,height=1.5in,clip,keepaspectratio]
{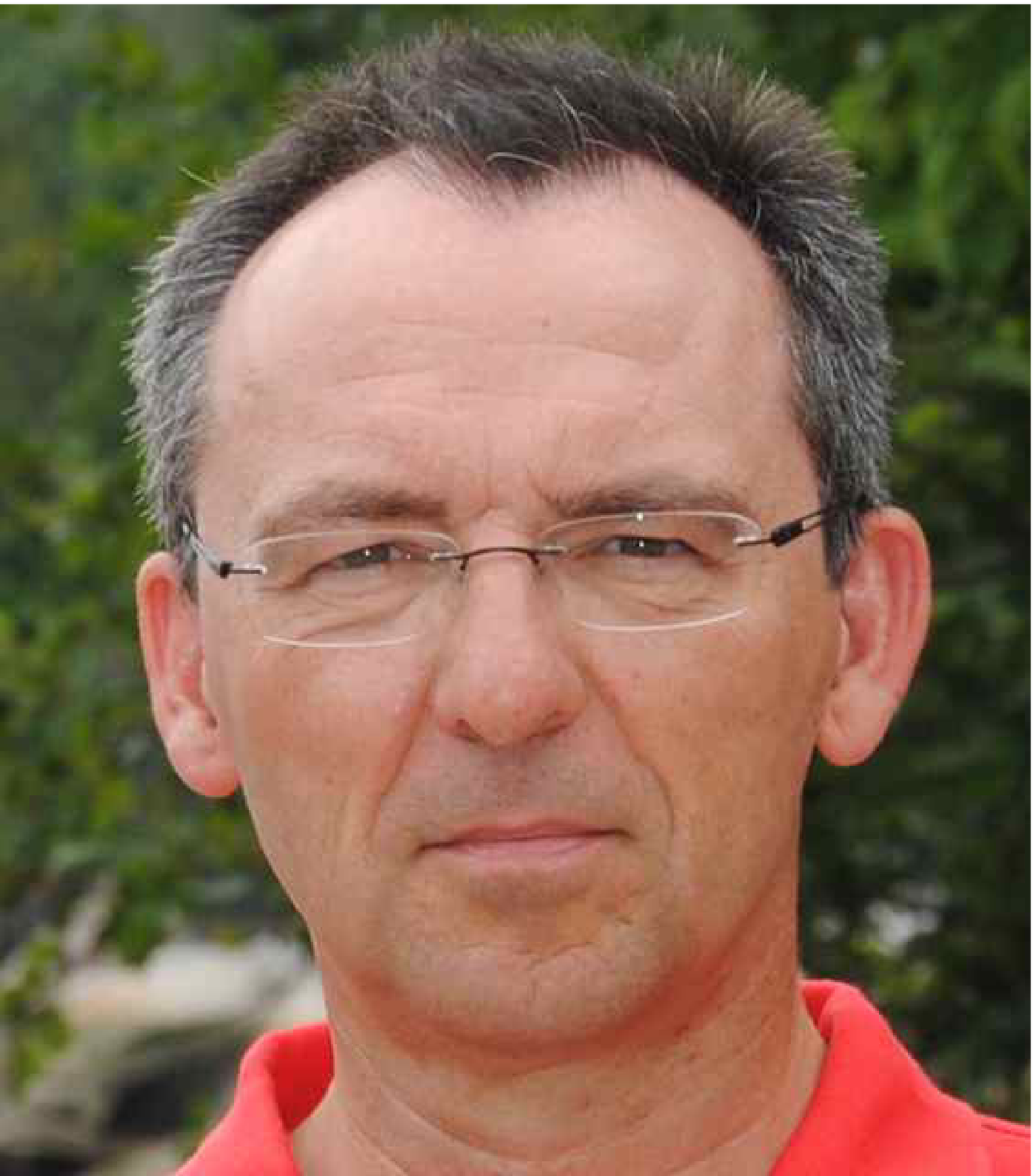}}]{Luc Vandendorpe}(F'06)
was born in Mouscron, Belgium in 1962. He received the Electrical Engineering degree ({\it summa cum laude})
and the Ph. D. degree from the Universit?Catholique de Louvain (UCL) Louvain-la-Neuve, Belgium
in 1985 and 1991 respectively. Since 1985, he is with the Communications and Remote
Sensing Laboratory of UCL where he first worked in the field of bit rate reduction techniques
for video coding. In 1992, he was a Visiting Scientist and Research Fellow at the Telecommunications
and Traffic Control Systems Group of the Delft Technical University, The Netherlands, where he worked
on Spread Spectrum Techniques for Personal Communications Systems.
From October 1992 to August 1997, L. Vandendorpe was Senior Research Associate of the Belgian
NSF at UCL, and invited assistant professor. Presently he is Professor and head of
the Institute for Information and Communication Technologies, Electronics and Applied Mathematics.

His current interest is in digital communication systems and more precisely resource
allocation for OFDM(A) based multicell systems, MIMO and distributed MIMO,
sensor networks, turbo-based communications systems, physical layer security and UWB based positioning.
In 1990, he was co-recipient of the Biennal Alcatel-Bell Award from the Belgian NSF
for a contribution in the field of image coding. In 2000 he was co-recipient
(with J. Louveaux and F. Deryck) of the Biennal Siemens Award from the Belgian NSF
for a contribution about filter bank based multicarrier transmission.
In 2004 he was co-winner (with J. Czyz) of the Face Authentication Competition,
FAC 2004. L. Vandendorpe is or has been TPC member for numerous IEEE conferences
(VTC Fall, Globecom Communications Theory Symposium, SPAWC, ICC) and for the Turbo Symposium.
He was co-technical chair (with P. Duhamel) for IEEE ICASSP 2006.
He was an editor of the IEEE Trans. on Communications for Synchronisation and
Equalization between 2000 and 2002, associate editor of the IEEE Trans. on
Wireless Communications between 2003 and 2005, and associate editor of the
IEEE Trans. on Signal Processing between 2004 and 2006. He was chair of the
IEEE Benelux joint chapter on Communications and Vehicular Technology
 between 1999 and 2003. He was an elected member of the Signal Processing for
Communications committee between 2000 and 2005, and an elected member of the Sensor
Array and Multichannel Signal Processing committee of the Signal Processing Society between 2006 and 2008.
He was an elected member of the Signal Processing for
Communications committee between 2000 and 2005,
and between 2009 and 2011, and an elected member
of the Sensor Array and Multichannel Signal Processing committee of the
Signal Processing Society between 2006 and 2008. He is the
Editor in Chief for the EURASIP Journal on Wireless Communications and
Networking and a Fellow of the IEEE.
\end{IEEEbiography}

\end{document}